\documentclass[prd, 10pt,aps,nofootinbib, floatfix, notitlepage,%twocolumn,
longbibliography,superscriptaddress]{revtex4-2}

\usepackage{amsmath,amsfonts,amssymb}
\usepackage[%colorlinks=true
citecolor=blue,urlcolor=blue]{hyperref}
\usepackage{graphicx}
\usepackage{mathrsfs}
\usepackage{xcolor}
\usepackage{cleveref}
\usepackage[export]{adjustbox}
\usepackage{enumitem}

\newcommand{\gal}{\hat{g}_{a\ell}}
\newcommand{\gae}{\hat{g}_{ae}}
\newcommand{\gamu}{\hat{g}_{a\mu}}
\newcommand{\galScalar}{g_{a\ell}}

\newcommand{\gamuScalar}{g_{a\mu}}
\newcommand{\gag}{g_{a\gamma}}
\newcommand{\gagEff}{g_{a\gamma}^{{\rm eff}}}
\newcommand{\gagDec}{g_{a \gamma}^{{\rm (D)}}}
\newcommand{\gagPri}{g_{a\gamma}^{{\rm (P)}}}
\newcommand{\diff}{\mathrm{d}}

\begin{document}
	
\title{ALP couplings to muons and electrons:\texorpdfstring{\\}{}
        a comprehensive analysis of supernova bounds}

\author{Ricardo~Z.~Ferreira}
\affiliation{Centro de F\'isica das Universidades do Minho e do Porto (CF-UM-UP),\\
Universidade do Minho, P-4710-057 Braga, Portugal}
\affiliation{Faculdade de Ci\^{e}ncias e Tecnologia and CFisUC, Departamento de F\'isica,\\ 
Universidade de Coimbra, Rua Larga P-3004-516 Coimbra, Portugal}
\author{M.C.~David~Marsh}
\affiliation{The Oskar Klein Centre for Cosmoparticle Physics,
Department of Physics,
Stockholm University, AlbaNova, 10691 Stockholm, Sweden}
\author{Eike~Ravensburg}
\affiliation{The Oskar Klein Centre for Cosmoparticle Physics,
Department of Physics,
Stockholm University, AlbaNova, 10691 Stockholm, Sweden}
\affiliation{CP3-Origins, University of Southern Denmark
Campusvej 55, 5230 Odense M, Denmark}

\begin{abstract}
    We provide a comprehensive analysis of the phenomenology of axion-like particles (ALPs) produced in core-collapse supernovae (ccSNe) through interactions with electrons and muons, both of which have a non-negligible abundance in the SN plasma. We identify and calculate six significant ALP-production channels, two of which are loop-level processes involving photons. We then examine several observational constraints on the ALP-electron and ALP-muon parameter spaces. Those include the bounds on anomalous cooling, energy deposition, decay into photons, diffuse gamma rays, and the 511 keV line. 
    Our results provide updated and robust constraints on ALP couplings to electrons and muons from an improved treatment of production and absorption processes. Furthermore, we quantify the uncertainties of the results by using three state-of-the-art supernova models based on two independent simulation codes, finding that constraints vary by  factors of ${\cal O}(2$--$10)$.
\end{abstract}

\maketitle

\tableofcontents

\section{Introduction}

Hot and dense core-collapse supernovae (ccSNe) are known as efficient factories of sub-GeV particles, and can thus be used as laboratories to study new physics. In recent years, much progress has been made in this direction, in particular in the case of Axion-Like Particles (ALPs), building upon the extensive literature on neutrinos from SNe \cite{Falk:1978kf,Raffelt:1996wa,
Beacom:2010kk}. 

ALPs can be efficiently produced in ccSNe if they couple to the Standard Model particles that exist in high densities at the core. The resulting ALP spectrum can then been systematically constrained by several observational considerations. The observed duration of the neutrino burst from the most recent nearby supernova, SN 1987A, constrains the energy emitted through ALPs. Those ALPs can escape the proto-neutron star at the center of the explosion, and lead to a rapid cooling \cite{Raffelt:1990yz}. The absence of a follow-up gamma-ray excess after the SN 1987A explosion limits the number of ALPs produced in the supernova that subsequently decay into photons outside the progenitor star \cite{Jaeckel:2017tud,Hoof:2022xbe,Ferreira:2022xlw,Muller:2023vjm}. The observation that SNe have a typical explosion energy \cite{Bruenn:2014qea,Nomoto:2013oal} places a limit on the amount of energy that can be dumped into the exploding stellar material from ALPs produced in the proto-neutron star \cite{Falk:1978kf,Caputo:2022mah} --- the so-called explosion energy bound.
Finally, the expected cosmological and galactic population of ccSNe provides a cumulative ALP spectrum that can contribute to the diffuse gamma-ray background \cite{Calore:2020tjw,Calore:2021klc,Caputo:2021rux,Muller:2023vjm}, if ALPs are very massive and decay to photons, and to the 511 keV line flux, if the decay happens to electrons and positrons \cite{Calore:2021klc,DelaTorreLuque:2023huu}.

Recently, progress has been made to extend the analysis to ALPs that couple to electrons \cite{Lucente:2021hbp,Ferreira:2022xlw,Fiorillo:2025sln} and muons \cite{Bollig:2020xdr,Croon:2020lrf,Caputo:2021rux}.
In Ref.~\cite{Ferreira:2022xlw}, we explored the case of an ALP derivatively coupled to electrons, so that the coupling to photons at low-energies is zero at tree-level and one can isolate the effect of the ALP-electron coupling separately. Our work included ALP production via bremsstrahlung, electron-positron fusion and two novel loop-induced processes, Primakoff and Photon coalescence. We showed that the latter, which are quite important for dark matter searches for ALPs coupled to electrons \cite{Ferreira:2022egk}, also have a prominent role of opening a new detection channel, the decay of ALPs to photons, that allowed to probe new regions in parameter space.

In this work, we provide a comprehensive analysis that extends previous works at the level of production and absorption processes and observational constraints of ALPs only coupled to leptons at tree-level (see also Ref.~\cite{Ravensburg:2023uin} for preliminary work in that direction). This work also provides an important estimation of the uncertainties of the results on the modelling of the ccSNe,  by comparing the predictions from distinct supernova simulations. 
As new processes, we will now include the ALP production and absorption via (semi-)Compton scatterings and lepton-pair annihilation that have been neglected in previous work \cite{Lucente:2021hbp,Ferreira:2022xlw} (see, however, the recent Ref.~\cite{Fiorillo:2025sln}). The former we now show to be important for a wide range of ALP masses. We then perform the same complete study for ALPs derivatively coupled to muons. We calculate and combine all these contributions for ALP-electron and ALP-muon interactions, thus completing previous studies that only included some of the processes and/or used approximated rates \cite{Bollig:2020xdr,Croon:2020lrf,Caputo:2021rux,Ferreira:2022xlw,Fiorillo:2025sln}.

Regarding observational constraints, we improve on our work in \cite{Ferreira:2022xlw} in multiple ways. We apply the explosion energy bound to ALP-electron couplings and improve over a previous analysis done for the ALP-muon coupling \cite{Caputo:2021rux}. 
Moreover, we revisit the computation of the diffuse gamma-ray spectrum and  provide a more general formula, based on similar phenomena for neutrinos \cite{Fogli:2004gy}, that is valid beyond the ultrarelativistic limit that has been used in previous literature \cite{Caputo:2021rux}.
We calculate the $511$ keV line flux coming from the decays of ALPs into electron-positron pairs in our galaxy and use the data from the SPI instrument  \cite{Bouchet:2010dj,Siegert:2015knp} to place constraints on the ALP-lepton coupling, following \cite{Calore:2021klc,DelaTorreLuque:2023huu}. Furthermore, we also reevaluate the bounds derived from cooling and from the absence of a gamma-ray excess at the time of SN~1987A explosion.   
At last, we highlight that we perform the analysis separately for three ccSNe models, from two independent codes \cite{Fischer:2021jfm,Bollig:2020xdr}. 
We believe this to be an important step in assessing the systematic uncertainties that underlay these studies and the derived constraints on the fundamental parameters.

This paper is organized as follows. In \cref{sec:EFT}, we discuss the ALP-lepton couplings we use in this work and describe the effective loop-induced photon ALP-photon coupling. In \cref{sec:SNmodel}, we describe the SN models that we use. For a  ``leptophilic'' ALP, up to six different production (and reabsorption) channels can yield a significant number of ALPs, of which two are loop-level processes, as we will discuss in \cref{sec:production}. Having established the number and energy of produced ALPs, we examine five different observational constraints on their parameter space in \cref{sec:bounds}. We conclude in \cref{sec:conclusion}.

\section{ALP-lepton EFT}
\label{sec:EFT}
\subsection{Types of ALP-lepton couplings}
\label{subsec:leptonCouplings}
ALPs can couple to the Standard Model (SM) in several ways, see e.g.~Refs.~\cite{Bauer:2020jbp,Chala:2020wvs,DiLuzio:2020wdo}.
Here, we focus on the flavor-diagonal interaction between ALPs and either of the two lightest leptons, the electron and the muon, both of which are present in the plasma of SN explosions. 
In order to isolate the effect of the ALP-lepton coupling we follow the approach in \cite{Ferreira:2022xlw} and consider models without tree-level ALP-photon interactions, i.e., a term like $\frac{\gag}{4} a \, F_{\mu\nu} \tilde F^{\mu\nu}$, is not present. There are UV-complete models that result in such an effective field theory (EFT) (so-called `photophobic' ALP models, see e.g.~Ref.~\cite{Craig:2018kne}).

Since the EFT for a massless ALP should be invariant under a constant shift of the ALP field $ a \to a + {\rm const.} $,
the only allowed dimension-five operator that couples a single lepton $\ell$ and the ALP is the derivative interaction \cite{Bonilla:2021ufe}
\begin{equation} \label{eq:alpLeptonInteractionLagrangian}
    \mathcal{L}_{a\ell} \supset 
    \gal (\partial_\mu a) \, \bar\ell \gamma^\mu \gamma^5 \ell \, ,
\end{equation}
where we take $\ell \in \{e, \mu\}$.

The often used pseudoscalar interaction term $ \mathcal{L}_{a\ell}^{\rm (PS)} = -i \galScalar \, a \, \bar\ell \gamma^5 \ell $ 
is related to our Lagrangian through a redefinition of the lepton field $\ell \to \exp\left( i \gal a \gamma_5 \right) \ell$ yielding
\begin{equation} \label{eq:alpLeptonPseudoScalarInteractionLagrangian}
\begin{aligned}
    \mathcal{L}_{a\ell} &\to
    - m_\ell \, \bar\ell \exp\left(2 i \gal \, a \gamma_5\right) \ell
    + \frac{\alpha}{2\pi} \gal \, a \, F_{\mu\nu} \tilde F^{\mu\nu}\\
    &\simeq -i \underbrace{2 m_\ell \gal}_{\equiv \galScalar} a \, \bar\ell \gamma^5 \ell
    + \frac{\alpha}{2\pi} \gal \, a \, F_{\mu\nu} \tilde F^{\mu\nu} \, ,
\end{aligned}
\end{equation}
with the lepton mass in vacuum $m_\ell$ and where, in the second line, we have omitted terms of order $\gal^2 a^2$ and higher as well as the fermion mass term. Here, the first term is due to the transformed fermion mass term and the second term has to be included because the transformation corresponds to an anomalous symmetry that does not leave the fermion path integral measure invariant --- see, e.g., refs.~\cite{Eberhart:2025lyu,Quevillon:2021sfz,Grojean:2023tsd} for more details.

We note that while $\mathcal{L}_{a\ell}$ is manifestly shift invariant, the same does not hold for the truncated $\mathcal{L}_{a\ell}^{\rm (PS)}$. The transformation above shows that this is because the pesudoscalar interaction Lagrangian is only the first term in an infinite tower of ALP-fermion interactions. Summing them yields the exponential in \cref{eq:alpLeptonPseudoScalarInteractionLagrangian} which again is shift invariant. A common UV completion of the effective ALP Lagrangian that is of the form $\mathcal{L}_{a\ell}^{\rm (PS)}$ is the DFSZ-like ALP \cite{Zhitnitsky:1980tq,Dine:1981rt,Arias-Aragon:2022iwl}. In our approach, this model would correspond to the derivative interaction in \cref{eq:alpLeptonInteractionLagrangian} plus an additional ``tree-level'' ALP-photon interaction $\gag = -\frac{2\alpha}{\pi} \gal$ that cancels the anomaly term in the transformed Lagrangian.

Since the field redefinition corresponds to just a change of variables in the path integral, the two Lagrangians in \cref{eq:alpLeptonInteractionLagrangian,eq:alpLeptonPseudoScalarInteractionLagrangian} yield the same physical observables \cite{Chisholm:1961tha,Kamefuchi:1961sb}. Thus, for processes only involving one ALP and the (anti-)lepton fields, the derivative and pseudoscalar interaction terms predict the same scattering cross section. On the other hand, the ``tree-level'' photon interaction (note that it is actually of third order in coupling constants) in the transformed Lagrangian affects the effective coupling to photons that we will explore in the next subsection.

\subsection{Effective ALP-photon coupling}
\label{subsec:effectivePhotonCoupling}
Models where the only tree-level interactions are between ALPs and charged leptons, can still feature ALP interaction with  photons through quantum loops. Subtly, this is the case even though the ALP-photon coupling is not generated through RGE running by the ALP-lepton interaction \cite{Chala:2020wvs,Bauer:2020jbp}.
The fully off-shell effective photon coupling can be written as \cite{Quevillon:2019zrd,Ferreira:2022xlw,Eberhart:2025lyu}
\begin{equation} \label{eq:gagEffOffShell}
    \gagEff(q_1^2,q_2^2, q_a^2) = \frac{2 \alpha}{\pi} \gal \left[ 1 + 2 m_\ell^2 C_0\left( q_1^2, q_2^2, q_a^2, m_\ell^2, m_\ell^2, m_\ell^2 \right) \right] \, ,
\end{equation}
where $q_{1,2}$ are the photon 4-momenta, $q_a$ is the ALP 4-momentum, $\alpha$ is the QED fine-structure constant, and $C_0$ is the three-point scalar Passarino-Veltman function following the definitions of Ref.~\cite{Shtabovenko:2016sxi}. Below, we will use this effective coupling for on-shell ALPs, $q_a^2 = m_a^2$ with the ALP mass $m_a$, and either one or both photons on-shell. Both, ALP-production and ALP-reabsorption processes in the SN plasma, as well as ALP decays, are facilitated in this way, and can influence potentially detectable signals. We note that the effective coupling vanishes in the $m_\ell \to \infty$ limit. For a pseudoscalar coupling, the effective photon interaction does not have the first, constant term in the square brackets in \cref{eq:gagEffOffShell}, and thus becomes non-vanishing in the heavy lepton-mass limit --- contrary to the generic behavior of higher-dimensional operators in EFTs.

\section{Supernova Models}
\label{sec:SNmodel}
\begin{figure}[b]
    \centering
    \includegraphics[width=.99\linewidth]{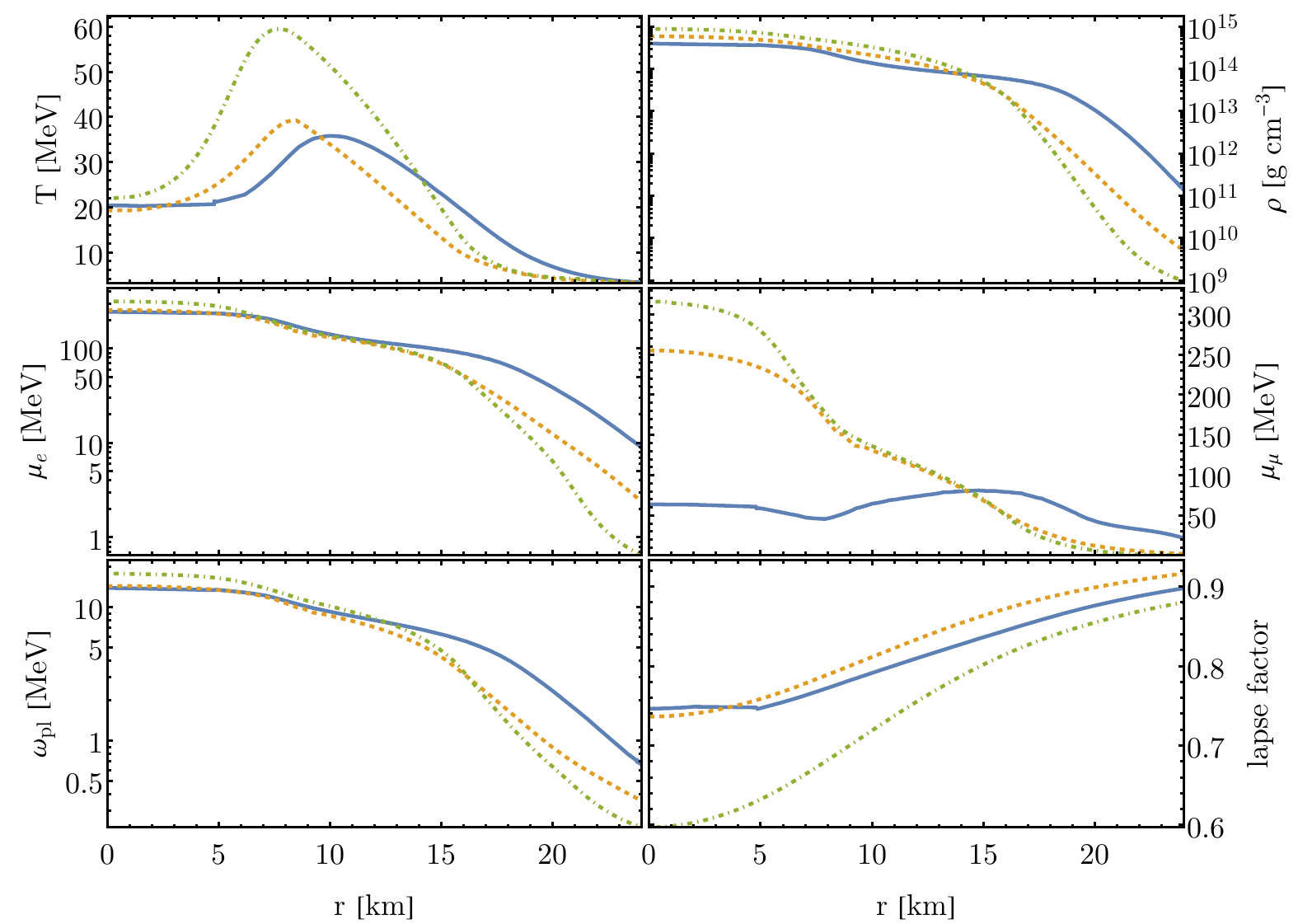}
    \caption{Snapshots at $t=1$~s post bounce of plasma properties for the three SN models; from left to right and top to bottom: the temperature, the mass density, the electron chemical potential, the muon chemical potential, the plasma frequency, and the gravitational lapse factor. The solid blue lines correspond to the DD2-18.0 model, the dashed orange lines to the SFHo-18.8 model, and the dot-dashed green lines to the LS220-20.0 model (see main text for details).}
    \label{fig:SNmodel}
\end{figure}

To calculate the number and energy of ALPs produced in a ccSN core we post-process models of the plasma properties that are obtained by hydrodynamical simulations. We select three spherically symmetric, i.e., one-dimensional, models that include muons and their interactions and which are compatible with the observed properties of SN~1987A:

\begin{description}[labelwidth=3cm,leftmargin=*,style=nextline]
  \item[DD2-18.0] First, as in our earlier work \cite{Ferreira:2022xlw}, we use the reference model of Ref.~\cite{Fischer:2021jfm} calculated by the AGILE-BOLTZTRAN simulation code \cite{Mezzacappa:1993gn,Liebendoerfer:2002xn}.
This model, which we refer to as DD2-18.0, 
assumes an $18 M_{\odot}$ progenitor star, the nuclear equation of state \textit{DD2} \cite{Fischer:2013eka}, and results in a final neutron star mass of $1.54 M_{\odot}$.
  \item[SFHo-18.8] Second, we use the coldest model of Ref.~\cite{Bollig:2020xdr}, which we refer to as SFHo-18.8.
This model was developed using  the Prometheus-Vertex code and is accessible via the Garching Core-Collapse Supernova Archive \cite{garchingArchive}. Starting from progenitor masses of $18.8 M_{\odot}$, using the \textit{SFHo} nuclear equation of state, this model results in   a final neutron star of mass $1.35 M_{\odot}$. This value is near the smallest value consistent with the remnant of SN~1987A \cite{Page:2020gsx}.
  \item[LS220-20.0] Finally, we consider the hottest model of Ref.~\cite{Bollig:2020xdr}, which was obtained  starting from a  $20.0 M_{\odot}$ progenitor star, using the \textit{LS220} nuclear equation of state, ending a final neutron star of mass of $ 1.93 M_{\odot}$ \cite{garchingArchive}. This large mass is above the largest expected remnant mass of SN~1987A \cite{Page:2020gsx}. Moreover, it should be noted that the \textit{SFHo} nuclear equation of state is more favored by theory, experiments, and observations as a description of the matter in the proto-neutron star, compared to \textit{LS220} \cite{Page:2020gsx}.
\end{description}

We choose this set of models to represent two different codes and three different equations of state, resulting in considerably different profiles and peak temperatures in the plasma. All subsequent calculations of this paper have been performed for each of these models, and the resulting differences in the outcomes highlight their sensitivity to some of uncertainties in the modeling of core-collapse SNe. 
In \cref{fig:SNmodel} we show the profiles for the temperature, energy density, electron and muon chemical potential, plasma frequency, and the gravitational lapse factor (encoding the redshift a particle emitted at that radius would experience when escaping to infinity) around the supernova core for the aforementioned models. While DD2-18.0 and SFHo-18.8 have comparable peak temperatures between $35$ and $40$~MeV, the plasma reaches $60$~MeV for LS220-20.0, and a similar hierarchy is observed for the maxima of the remaining four quantities. The muon chemical potential is a particularly extreme case, since in the DD2-18.0 model it is a factor of $\sim 5$ smaller at its peak than in the other two models. The lapse factor is smallest for LS220-20.0, since gravitational effects are most pronounced in that model. Notably, the other shown quantities fall off less quickly at large radii for the DD2-18.0 model than for the two Garching models. One effect of that, as we will see, is that overall more ALPs are produced in that model than for SFHo-18.8, even though its peak temperature is comparable and its peak density is slightly lower.

Note that the aforementioned simulations do not include the presence of ALPs, which could back-react on the dynamics of the plasma if they are produced copiously enough. Our approach is perturbative in the sense that the weakly interacting ALPs are only considered as a small perturbation in the plasma that does not alter the explosion dynamics in a significant way. We will discuss potential limitations of this approach, namely when ALPs become relatively strongly coupled to the plasma, in \cref{subsec:calorimetricBound}.

\section{ALP production in the supernova core}
\label{sec:production}

In this section, we calculate the ALP production spectrum, the number of ALPs produced per differential of volume, time, and energy. This is the basic quantity for the determination of any ALP-induced signal or constraint and fully characterizes the isotropic production of ALPs in a SN explosion. From the collision term in the Boltzmann equation, we can derive the following expression for the production spectrum \cite{Ferreira:2022xlw,Raffelt:1990yz}:
\begin{equation}
\begin{split} \label{eq:spectrumDefinition}
    \frac{\diff^2 n_a}{\diff t \, \diff\omega_a} =
    \Bigg[ &\prod_i \int \frac{\diff^3 \vec{p}_i}{(2\pi)^3 2 E_i} f_i(E_i) \Bigg]
    \Bigg[ \prod_{j \neq a} \int \frac{\diff^3 \vec{p}\,'_j}{(2\pi)^3 2 E'_j} \left[ 1 \pm f_j(E'_j) \right] \Bigg]  (2\pi)^4 \delta^{(4)} \Bigg( \sum_i p_i - \sum_j p'_j \Bigg) S \,\,
    \frac{\lvert\vec{p}\,'_a\rvert}{4\pi^2} \lvert \mathcal{M} \rvert^2 \, ,
\end{split}
\end{equation}
where $ p_i = (E_i, \, \vec{p}_i) $ are the incoming particles' 4-momenta, energies, and 3-momenta, respectively, $ p'_j = (E'_j, \, \vec{p}\,'_j) $ those of the outgoing particles including the ALP (for which we set $ E_a \equiv \omega_a $), $ f_{i,j} $ are the respective phase-space distribution functions, $ |\mathcal{M}|^2 $ is the total squared matrix element for the respective ALP production process summed over initial and final state polarisations, and $ S = 1/n! $ is a symmetry factor to avoid overcounting of $ n $ identical particles in the initial or the final state. The quantum statistical factor for a final state particle is $ 1 + f_j $ if $ j $ represents a boson, and $ 1 - f_j $ for a fermion. Note that the product over final state momenta does not include the ALP's phase space as $ \frac{\diff^2 n_a}{\diff t \, \diff\omega_a} $ is a differential with respect to the ALP's energy in the plasma frame; the difference between the differentials of ALP energy and Lorentz-invariant phase space yields the factor $ \frac{\lvert\vec{p}'_a\rvert}{4\pi^2} $. Finally, we have neglected a factor $1+f_a$, assuming that the ALP phase space density is negligible. This assumption is only violated for relatively strong couplings, when the ALPs interact frequently with the plasma and the number and energy of ALPs that are transferred out of the SN is mostly dictated by the ALP's mean-free path, which is independent of the ALP phase-space density $f_a$ (see also \cref{footnote:boseEnhancementProdSpectrum} in \cref{subsec:cooling}).

Here and in the following we define
\begin{equation} \label{eq:distributionFunctionDefinitions}
    f_{\rm F}^{\pm}(E) = \frac{1}{\exp\left( \frac{E \pm \mu}{T} \right) + 1} \, , \quad
    f_{\rm B}(\omega) = \frac{1}{\exp\left( \frac{\omega}{T} \right) - 1} \, ,
\end{equation}
for the Fermi-Dirac distribution function of (positively) negatively charged (anti-)leptons with chemical potential $\pm\mu$ where $\mu>0$, and the Bose-Einstein distribution function of photons.

\begin{figure}[h]
    \includegraphics[width=.28\textwidth, valign=t]{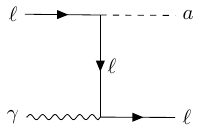}
    \hspace{.01\textwidth}
    \includegraphics[width=.28\textwidth, valign=t]{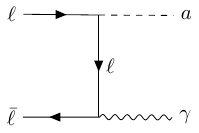}
    \hspace{.01\textwidth}
    \includegraphics[width=.28\textwidth, valign=t, raise=-.04cm]{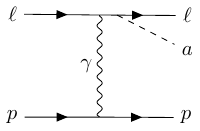}
    \\[.5cm]
    \hspace{1.cm}
    \includegraphics[height=4cm, valign=c]{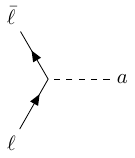}
    \hspace{.5cm}
    \includegraphics[width=.28\textwidth, valign=c]{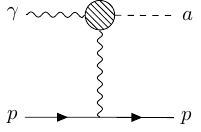}
    \hspace{1.5cm}
    \includegraphics[height=4cm, valign=c, raise=.05cm]{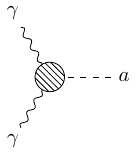}
    \caption{Processes contributing to ALP production and reabsorption in the SN plasma. From top left to bottom right: Compton scattering (\labelcref{subsec:compton}), lepton-pair annihilation (\labelcref{subsec:annihilation}), Bremsstrahlung (\labelcref{subsec:bremsstrahlung}), electron-positron fusion (\labelcref{subsec:fusion}), Primakoff process (\labelcref{subsec:primakoff}), and photon coalescence (\labelcref{subsec:coalescence}). The blobs in the latter two diagrams indicate the effective, loop-induced ALP-photon coupling. Note that crossed diagrams are not shown.}
    \label{fig:feynmanDiagrams}
\end{figure}

The SN model enters the production spectra through the thermal distribution functions $f_{i,j}$ but also through thermal modifications to the matrix elements such as effective masses or screening scales. We will discuss the relevant effects for each process separately. In the following \cref{subsec:compton,subsec:annihilation,subsec:bremsstrahlung,subsec:fusion,subsec:primakoff,subsec:coalescence}, we will discuss the matrix elements for the relevant processes, shown in \cref{fig:feynmanDiagrams}, and calculate the resulting production spectra. The first two, Compton production of ALPs and lepton-pair annihilation, were not included in Ref.~\cite{Ferreira:2022xlw}, while other analyses \cite{Lucente:2021hbp,Fiorillo:2025sln,Bollig:2020xdr} neglected the loop-induced Primakoff and photon coalescence processes.

    \subsection{Compton} \label{subsec:compton}
    It has been argued \cite{Pantziris:1986dc,Raffelt:1996wa,Lucente:2021hbp} that electron-nucleon Bremsstrahlung and electron-positron fusion are the only relevant production processes for ALPs coupled to electrons in ccSNe. We recently pointed out \cite{Ferreira:2022xlw} that the Primakoff and photon coalescence process, which are possible at the loop level even if $\gag = 0$, do contribute a non-negligible amount of ALPs. Here we show that there is yet another \emph{tree-level} process that yields a very important contribution to the ALP production spectrum in ccSNe: the (semi-)Compton or photoproduction process $\ell^{-} \, \gamma \to a \, \ell^{-} $, shown in the first panel of \cref{fig:feynmanDiagrams}. Ref.~\cite{Caputo:2021rux} recently argued, based on approximated emission rates in the massless ALP limit, that the Compton process is the dominant production process in the case of muons coupled to ALPs via a pseudoscalar interaction.
    We expand their arguments by using exact expressions to calculate the Compton production spectrum for ALP-electron interactions and all production processes induced by the ALP-muon interaction, and by including the ALP mass corrections (cf.~\cite{Fiorillo:2025sln} for related recent work).

    We calculate the matrix element for the Compton process in terms of the Mandelstam variables $ s, t, u $ and eliminate $ s $ from the resulting expression using $ s = 2m_\ell^2 + m_a^2 + m_\gamma^2 - t - u $. Here, $m_\ell$ is the lepton mass, $m_a$ the ALP mass, and $m_\gamma$ the effective photon mass in the plasma, which in the SN core is the plasma frequency \cite{Kopf:1997mv}
    \begin{equation}
        m_\gamma \simeq 16.3~{\rm MeV} \, Y_e^{1/3} \left( \frac{\rho}{10^{14} \, {\rm g \, cm}^{-3}} \right)^{1/3} \, ,
    \end{equation}
    where $Y_e(r,t)$ is the fraction of the number of electrons in the plasma and $\rho(r,t)$ is the mass density, dominated by nucleons. Note that the effective mass is not identical to the plasma frequency but instead $m_\gamma^2= \frac{3}{2} \omega_{\rm pl}^2 $ (this corrects the corresponding equation in refs.~\cite{Lucente:2020whw,Ferreira:2022xlw}).
    Then, upon aligning the $z$-axis of the local plasma frame with the ALP 3-momentum $ \vec p_a $, we can express the matrix element easily as function of only the masses, three energies, and $ c_{\rm in}, \, c_\gamma $, the cosines of the two angles between the incoming lepton 3-momentum $ \vec{p}_\ell\,^{\rm in} $ and $ \vec p_a $, and between the photon 3-momentum $ \vec p_\gamma $ and $ \vec p_a $, respectively. This yields the expression
    \begin{equation} \label{eq:comptonMatrixElement}
    \begin{aligned}
        |\mathcal{M}|^2 &=
        4 e^2 \galScalar^2 \Bigg\{
        \frac{m_a^2 \left(2 m_\ell^2 + m_{\gamma}^2\right)}{\left[2 \omega_a \left(\omega_\gamma \left(1 - \beta_a \beta_\gamma c_\gamma\right) + E_\ell^{\rm in} \left(1 - \beta_a \beta_\ell^{\rm in} c_{\rm in}\right)\right) - m_a^2 \right]^2}\\
        &+ \frac{m_a^2 \left(2 m_\ell^2 + m_{\gamma}^2 + m_a^2\right) - 4 \omega_a^2 (E_\ell^{\rm in})^2 \left(1 - \beta_a \beta_\ell^{\rm in} c_{\rm in}\right)^2}{\left[2 \omega_a E_\ell^{\rm in} \left(1 - \beta_a \beta_\ell^{\rm in} c_{\rm in}\right) - m_a^2\right]^2}
        + \frac{2 \omega_a \omega_{\gamma} \left(1 - \beta_a\beta_{\gamma} c_{\gamma}\right)}{2 \omega_a E_\ell^{\rm in} \left(1 - \beta_a \beta_\ell^{\rm in} c_{\rm in}\right) - m_a^2}\\
        &+ \frac{4 \omega_a^2 (E_\ell^{\rm in})^2 \left(1 - \beta_a \beta_\ell^{\rm in} c_{\rm in}\right)^2 - m_a^2 (4 m_\ell^2 - m_a^2)}{\left[2 \omega_a E_\ell^{\rm in} \left(1 - \beta_a \beta_\ell^{\rm in} c_{\rm in}\right) - m_a^2\right] \left[2 \omega_a \left(\omega_{\gamma} \left(1 - \beta_a \beta_{\gamma} c_{\gamma}\right) + E_\ell^{\rm in} \left(1 - \beta_a \beta_\ell^{\rm in} c_{\rm in}\right)\right) - m_a^2\right]}
        \Bigg\} \, ,
    \end{aligned}
    \end{equation}
    where $e$ is the QED coupling constant, the energies are $\omega_{a,\gamma} = \sqrt{p_{a,\gamma}^2 + m_{a,\gamma}^2} $ and $ E_{\rm in} = \sqrt{p_{\rm in}^2 + m_\ell^2} $, with $ p_x \equiv |\vec p_x| $, and $ \beta_x = p_x / \sqrt{p_x^2 + m_x^2} $ are the relativistic velocities. We have used the Mathematica package FeynCalc \cite{MERTIG1991345,Shtabovenko:2016sxi,Shtabovenko:2020gxv} to calculate the matrix element.
    Using \cref{eq:spectrumDefinition}, we find the production spectrum to be given by
    \begin{equation} \label{eq:comptonProdSpectrum}
    \begin{split}
        \frac{\diff n_a}{\diff t \, \diff\omega_a}\Big\rvert_{{\rm C}} =
        \frac{p_a}{256 \pi^6} \int_{m_\ell}^{\infty} \diff E_{\rm in} \int_{m_\gamma}^{\infty} \diff \omega_\gamma \int_{-1}^{1} \diff c_{\rm in} \int_{-1}^{1} \diff c_\gamma \, f_{\rm F}^{-}(E_{\rm in}) f_{\rm B}(\omega_\gamma) \left[1 - f_{\rm F}^{-}(E_{\rm in}+\omega_\gamma-\omega_a)\right]\\
        \times \frac{\Theta(1 - c_\delta^2)}{\sqrt{(1-c_\gamma^2)(1-c_{\rm in}^2)(1-c_\delta^2)}} \Theta(E_{\rm in} + \omega_\gamma - \omega_a - m_\ell) \, |\mathcal{M}|^2 \, ,
    \end{split}
    \end{equation}
    where $\Theta$ is the Heaviside function and energy conservation fixes $ c_\delta $, the cosine of the difference between the azimuthal angles of $\vec{p}_{\rm in}$ and $\vec{p}_\gamma$, to
    \begin{equation} \label{eq:cDeltaCompton}
        c_\delta = \frac{m_a^2 + m_\gamma^2 - 2 \omega_a E_{\rm in}(1 - \beta_a \beta_{\rm in} c_{\rm in}) - 2 \omega_a\omega_\gamma(1 - \beta_a \beta_\gamma c_\gamma) + 2 E_{\rm in} \omega_\gamma(1 - \beta_{\rm in}\beta_{\rm \gamma} c_{\rm in} c_\gamma)}{2 p_{\rm in} p_\gamma \sqrt{(1 - c_{\rm in}^2)(1 - c_\gamma^2)}} \, .
    \end{equation}
    We evaluate the integrals in \cref{eq:comptonProdSpectrum} numerically using the Suave algorithm for multidimensional integrals \cite{Hahn:2004fe}. To the best of our knowledge, this is the first time that the relativistic Compton-production spectrum for massive ALPs has been calculated without further approximations.

    Importantly, in all the above expressions (and all the following ones), the mass of on-shell electrons and positrons as well as the mass in the electron propagator, $m_e$, is the \emph{effective thermal} mass in an ultrarelativistic plasma \cite{Braaten:1991hg,Lucente:2021hbp}
    \begin{equation} \label{eq:effectiveLeptonMass}
        m_e = \frac{m_e^0}{\sqrt{2}} + \sqrt{\frac{\left(m_e^0\right)^2}{2} + \frac{\alpha}{\pi}\left(\mu_e^2 + \pi^2 T^2\right)} \, ,
    \end{equation}
    with the vacuum mass of the electron $m_e^0 = 511$~keV, the fine-structure constant $ \alpha = \frac{e^2}{4\pi} $, the chemical potential $\mu_e(r,t)$, and temperature $T(r,t)$. Since muons have $m_\mu^0 = 106$~MeV they are a non-relativistic component in the SN plasma and we can neglect thermal corrections to their dispersion relation \cite{Braaten:1991hg}. As discussed in Ref.~\cite{Lucente:2021hbp}, all matrix elements only involving the tree-level interaction between ALP and a lepton should yield the same result if the interaction is written in the pseudoscalar or derivative forms, see \cref{eq:alpLeptonInteractionLagrangian}. However, the relation between the coupling constants in the two cases, $\galScalar = 2 m_l^0 \, \gal$, involves the vacuum mass $m_\ell^0$ \cite{Lucente:2021hbp}. As remarked in Ref.~\cite{Ferreira:2022xlw}, these two statements seem to contradict each other because the ratio of squared and polarization averaged matrix elements calculated with pseudoscalar and derivative coupling (neglecting thermal effects beyond an effective mass) yields $\left|\mathcal{M}^{\rm (PS)}\right|^2 / \left|\mathcal{M}^{\rm (Der)}\right|^2 = \galScalar^2 / (4 p_\ell^2 \gal^2) = \left( m_\ell^0 / m_\ell^{\rm th} \right)^2$ since the square of the on-shell lepton momentum $p_\ell$ should be the on-shell, i.e., thermal mass, and thus, the two couplings seemingly yield different matrix elements. This might indicate that thermal effects beyond an effective mass play a relevant role in electron and positron propagation in the SN plasma. We leave the clarification of these thermal field theory questions for future studies as they go beyond the scope of this work, and have calculated and written the matrix element in \cref{eq:comptonMatrixElement} (and of all tree-level processes in the following sections) in terms of the dimensionless parameter $\galScalar$.

    The anti-lepton number density is much smaller than that of the respective leptons since they have a chemical potential with the opposite sign, see \cref{eq:distributionFunctionDefinitions}, and hence we can neglect the contributions from $\ell^{+} \gamma \to a \ell^{+}$.
    
    In \cref{fig:processComparison} we show the contributions of all production processes considered here to the total number of ALPs produced in the free-streaming regime, i.e., the volume, time, and energy integral over \cref{eq:comptonProdSpectrum} with possible reabsorption of strongly-coupled ALPs not taken into account. The Compton contribution is shown in red in \cref{fig:processComparison} and is further discussed in comparison to the other contributions in \cref{subsec:comparison}.

    \subsection{Lepton pair annihilation} \label{subsec:annihilation}
    Another ALP-production process is lepton pair annihilation $\ell^{-}\ell^{+} \to \gamma \, a$ via a virtual lepton. Due to the effective photon coupling, this process could also proceed via a virtual photon (in the s-channel), but this loop-contribution competes directly with the tree-level matrix elements and hence we do not consider it further. As with the Compton process discussed above, electron-positron annihilation was argued to be negligible, see e.g.~Refs.~\cite{Pantziris:1986dc,Raffelt:1996wa,Lucente:2021hbp}, because of the anti-lepton in the initial state, which as stated above has a very small phase-space density, while otherwise being comparable to the Compton process which does not suffer from such a suppression. Here, after already having set up the right frame to easily include an extra contribution, we check this estimate and find it mainly to be true in the free-streaming regime except for the ALP-muon coupling and ALP masses around 100~MeV where we observe that this channel could be play an important role.

    Using notation analogous to that in \cref{subsec:compton}, the matrix element of the annihilation process is:
    \begin{equation}
    \begin{aligned}
        |\mathcal{M}|^2 &= \frac{4 e^2 \galScalar^2}{\left(m_\ell^2 - t\right)^2 \
        \left(m_\ell^2-u\right)^2}
        \Bigg\{
        \left(m_\ell^2-t\right) \left(m_\ell^2-u\right) \left[ 2 m_a^4 + \left(t + u - 2 m_\ell^2\right)^2\right]\\
        &\qquad -m_a^2 \left[2 m_\ell^4 \left(m_{\gamma }^2-t-u\right)-2 m_\ell^2 \left(m_{\gamma }^2 (t+u)+2 t u\right) + 4 m_\ell^6 + m_\gamma^2 \left(t^2+u^2\right)+2 t u (t+u)\right]
        \Bigg\} \, ,
    \end{aligned}
    \end{equation}
    where the Mandelstam variables $t$ and $u$ can be expressed as
    \begin{equation}
    \begin{aligned}
        t &= m_\ell^2 + m_a^2 - 2 (E_\ell \, \omega_a - p_\ell \, p_a c_\ell) \, ,\\
        u &= m_\ell^2 + m_a^2 - 2 (E_{\bar\ell} \, \omega_a - p_{\bar\ell} \, p_a c_{\bar\ell}) \, ,
    \end{aligned}
    \end{equation}
    with $c_\ell, \, c_{\bar\ell}$ the cosine of the angle between lepton and ALP three-momenta, and anti-lepton and ALP three-momenta, respectively.
    This matrix element is consistent with the total cross section for lepton-annihilation calculated in Ref.~\cite{Mikaelian:1978jg} in the limit $m_\gamma \to 0$. Following the procedure laid out in \cref{subsec:compton}, we find for the production spectrum:
    \begin{equation} \label{eq:annihilationProdSpectrum}
    \begin{split}
        \frac{\diff n_a}{\diff t \, \diff\omega_a}\Big\rvert_{{\rm A}} =
        \frac{p_a}{256 \pi^6} \int_{m_\ell}^{\infty} \diff E_{\ell} \int_{m_\ell}^{\infty} \diff E_{\bar\ell} \int_{-1}^{1} \diff c_\ell \int_{-1}^{1} \diff c_{\bar\ell} \, f_{\rm F}^{-}(E_\ell) f_{\rm F}^{+}(E_{\bar\ell}) \left[1 + f_{\rm B}(E_\ell + E_{\bar\ell} - \omega_a)\right]\\
        \times \frac{\Theta(1 - c_\delta^2)}{\sqrt{(1-c_\ell^2)(1-c_{\bar\ell}^2)(1-c_\delta^2)}} \Theta(E_\ell + E_{\bar\ell} - \omega_a - m_\gamma) \, |\mathcal{M}|^2 \, ,
    \end{split}
    \end{equation}
    where again $c_\delta$ is the cosine of the difference between azimuthal angles of $\vec{p}_\ell$ and $\vec{p}_{\bar\ell}$:
    \begin{equation} \label{eq:cDeltaAnnihilation}
        c_\delta = \frac{2m_\ell^2 + m_a^2 - m_\gamma^2 - 2 \omega_a E_\ell(1 - \beta_a \beta_\ell c_\ell) - 2 \omega_a E_{\bar\ell}(1 - \beta_a \beta_{\bar\ell} c_{\bar\ell}) + 2 E_\ell E_{\bar\ell} (1 - \beta_\ell\beta_{\bar\ell} c_\ell c_{\bar\ell})}{2 p_\ell p_{\bar\ell} \sqrt{(1 - c_\ell^2)(1 - c_{\bar\ell}^2)}} \, .
    \end{equation}
    
    The resulting contribution from lepton annihilation to the total number of ALPs is shown in blue in \cref{fig:processComparison}.

    \subsection{Bremsstrahlung} \label{subsec:bremsstrahlung}
    The bremsstrahlung process $ \ell \, p \to \ell \, p \, a $, via the exchange of a virtual photon, is an important production process for ALPs in SNe, as was first shown in Ref.~\cite{Lucente:2021hbp}. Because the target proton is much heavier than the typical thermal energies of the incoming lepton, we can neglect the recoil of the proton such that energy conservation implies $E_{\rm in} \simeq \omega_a + E_{\rm out}$ with the incoming (outgoing) lepton energy $E_{\rm in\, (out)}$ and the ALP energy $\omega_a$. \Cref{eq:spectrumDefinition} yields the following expression for the production spectrum of ALPs created by lepton-proton Bremsstrahlung:
    \begin{equation}\label{eq:BremsstrahlungProduction}
        \frac{\diff^2 n_a}{\diff t \, \diff\omega_a}\Bigg\rvert_{\rm B} = \frac{n_p^{{\rm eff}}}{64\pi^6} \int_{-1}^1 \diff c_a \, \int_{-1}^1 \diff c_{\rm out} \, \int_{0}^{2\pi} \diff\delta \, \int_{m_e}^\infty \diff E_{\rm out} \, p_{\rm in} \, p_{\rm out} \, p_a \left\lvert \mathcal{M} \right\rvert^2 f_{\rm F}^-(E_{\rm in}) \left(1 - f_{\rm F}^-(E_{\rm out})\right) \, ,
    \end{equation}
    with the notation analogous to \cref{eq:comptonMatrixElement,eq:comptonProdSpectrum}; however, here we have chosen $\vec p_{\rm in}$ to be parallel to the $z$-axis, and hence $\delta$ is the angle between the planes spanned by $\vec p_a$ and $\vec p_{\rm in}$ and $\vec p_{\rm out}$ and $\vec p_{\rm in}$, respectively. Since the proton can provide momentum transfer, there is no momentum conservation between $\vec p_{\rm in}, \vec p_{\rm out}, \vec p_a$, and hence the (cosines of the) angles $c_{a, {\rm out}}$ and $ \delta$ are not restricted.
    This expression agrees with Ref.~\cite{Lucente:2021hbp}, was checked in Ref.~\cite{Ferreira:2022xlw}, and the full matrix element $|\mathcal{M}|^2$ can be found in Ref.~\cite{Carenza:2021osu}.
    
    We consider only free protons as targets since electrons are highly degenerate in the SN core, and ALP production is most efficient at high temperatures where the density of heavier nuclei can be neglected \cite{Lucente:2020whw}. Therefore, the number density of targets is the \emph{effective} number density of protons, taking their degeneracy into account \cite{Payez:2014xsa}
    \begin{equation}\label{eq:npEff}
        n_p^{{\rm eff}} = 2 \int \frac{\diff^3 p}{(2\pi)^3} f_p (1 - f_p) \, ,
    \end{equation}
    which can be up to $ \sim 60\% $ smaller than the naive number density, without the factor $ (1 - f_p) $, of free protons in SN~1987A (see also Ref.~\cite{Lucente:2020whw} for further discussion of this point in the DD2-18.0 model). The effective density $ n_p^{\rm eff} $ factors out of the expression for the spectrum in \cref{eq:BremsstrahlungProduction} in the no-recoil limit.
    As for the Compton process, we can neglect bremsstrahlung from anti-leptons due to their low number density.

    The bremsstrahlung contribution to the total ALP number produced in the SN is shown in green in \cref{fig:processComparison}.
    
    \subsection{Lepton fusion} \label{subsec:fusion}
    If the ALP is heavy enough, $m_a > 2 m_\ell$, where we remind the reader that $m_\ell$ is the effective thermal lepton mass shown in \cref{eq:effectiveLeptonMass}, lepton fusion $\ell^{-} \ell^{+} \to a$ provides a very efficient ALP production process. The lepton fusion production spectrum is \cite{Ferreira:2022xlw}
    \begin{equation}\label{eq:electronPositronFusionSpectrum}
        \frac{\diff^2 n_a}{\diff t \, \diff\omega_a}\Bigg\rvert_{F} = \frac{\galScalar^2 m_a^2}{16 \pi^3} \,\Theta(m_a - 2 m_\ell) \int_{E_{\rm min}}^{E_{\rm max}} \diff E_+ \, f_{\rm F}^-(\omega_a - E_+) f_{\rm F}^+(E_+) \, ,
    \end{equation}
    with the ALP energy $ \omega_a $, the anti-lepton energy $E_{+}$, energy conservation yielding $ \omega_a - E_{+} $ as the lepton energy, and momentum conservation leading to the upper and lower integration limits
    \begin{equation}\label{eq:eeFusionIntegrationLimits}
        E_{\rm max,min} = \frac{\omega_a}{2} \pm \frac{\sqrt{(\omega_a^2 - m_a^2)(m_a^2 - 4 m_\ell^2)}}{2 m_a} \, .
    \end{equation}

    Even though the anti-lepton number density is small, lepton fusion is a $2\to1$ process and involves $\galScalar$ as the only coupling constant, and thus is the leading production process for heavy ALPs, as shown in purple in \cref{fig:processComparison}.
    
    \subsection{Primakoff} \label{subsec:primakoff}
    Since the ALP-lepton interaction leads to a ALP-photon interaction at the one-loop level, ALPs can also be produced via the Primakoff process $\gamma \, p \to a \, p$ via a virtual photon exchange, as discussed in \cref{sec:EFT}. As for the bremsstrahlung process, protons are the only relevant targets \cite{Raffelt:1985nk}, and we can neglect their recoil. Thus, the production spectrum resulting from \cref{eq:spectrumDefinition} is \cite{Ferreira:2022xlw,DiLella:2000dn}
    \begin{equation}\label{eq:PrimakoffSpectrum}
        \frac{\diff^2 n_a}{\diff t \, \diff\omega_a}\Bigg\rvert_{P} = \frac{\alpha \, n_p^{{\rm eff}}}{4 \pi^2} \frac{p_\gamma^3 p_a^3 \, \Theta(\omega_a - m_\gamma)}{\exp(\omega_a/T) - 1} \int_{-1}^{1} \diff c_\theta \frac{\left\lvert \gagPri(\omega_a, c_\theta) \right\rvert^2 (1 - c_\theta^2)}{(p_\gamma^2 + p_a^2 - 2 p_\gamma p_a c_\theta)(p_\gamma^2 + p_a^2 - 2 p_\gamma p_a c_\theta + \kappa_{{\rm S}}^2)} \, ,
    \end{equation}
    where $c_\theta$ is the cosine of the angle between the photon and ALP three-momenta $\vec{p}_{\gamma,a}$, which in the no-recoil limit correspond to identical energies $\omega_a = \sqrt{p_a^2 + m_a^2} = \sqrt{p_\gamma^2 + m_\gamma^2}$, $\kappa_{\rm S} = e^2 n_p^{\rm eff} / T$ is a screening scale induced by plasma effects \cite{Raffelt:1985nk} (with the effective number of protons as in \cref{eq:npEff}), and the effective loop-induced coupling in the Primakoff process is \cite{Ferreira:2022xlw}
    \begin{equation} \label{eq:effectivePrimakoffCoupling}
    \begin{aligned}
        \gagPri(\omega_a, c_\theta) &= \frac{2 \alpha}{\pi} \gal \left[ 1 + 2 (m_\ell^0)^2 C_0\left( m_\gamma^2, t, m_a^2, (m_\ell^0)^2, (m_\ell^0)^2, (m_\ell^0)^2 \right) \right] \, , \quad \text{with } t = m_a^2 + m_\gamma^2 - 2 \omega_a^2 + p_a p_\gamma c_\theta \, , \\
        &\simeq
        \frac{2 \alpha}{\pi} \gal \times
        \begin{cases}
            - \frac{t + m_a^2 + m_\gamma^2}{12 (m_\ell^0)^2} & \text{for } \max(m_a, |t|, m_\gamma) \ll m_\ell^0 \, ,\\
            1 & \text{for } \max(m_a, |t|, m_\gamma) \gg m_\ell^0 \, ,
        \end{cases}
    \end{aligned}
    \end{equation}
    where $C_0$ is the scalar, three-point Passarino-Veltmann function \cite{Passarino:1978jh} defined as in Ref.~\cite{Shtabovenko:2016sxi}, which we evaluate numerically using LoopTools \cite{Hahn:1998yk}. The effective Primakoff coupling vanishes as the square of the largest relevant energy scale over the square of the lepton mass, if that scale is small compared to $m_\ell^0$, see Ref.~\cite{Ferreira:2022xlw,Eberhart:2025lyu} for more details. On the other hand, for a small lepton mass, the effective coupling becomes the constant $\frac{2 \alpha}{\pi} \gal$, which is mostly the relevant limit here, since the typical kinetic energy of particles near the center of the plasma is larger than $m_\ell^0$.
    
    Even though Primakoff production is a loop-level process, it cannot be neglected for ALPs coupled to electrons, as we can see from the Primakoff contribution to the total number of ALPs produced in the SN shown in orange in \cref{fig:processComparison}. For ALPs coupled to muons, the effective Primakoff coupling is suppressed because of the large mass of the muon. Additionally, the tree-level processes are enhanced by the larger muon mass, and thus, Primakoff production is negligible for the total ALP number produced in the explosion. However, the Primakoff process can still be a large contribution locally, e.g., at relatively large distances from the SN centre where the muon density rapidly declines, and hence all tree-level processes are suppressed. This is important whenever reabsorption of ALPs becomes relevant, see \cref{subsec:cooling,subsec:calorimetricBound}.

    \subsection{Photon coalescence} \label{subsec:coalescence}
    Similar to lepton fusion, at the one-loop level also photon coalescence $\gamma \, \gamma \to a$ can yield a contribution to ALP production, if $m_a > 2 m_\gamma$. With \cref{eq:spectrumDefinition} we find the production spectrum \cite{Ferreira:2022xlw}
    \begin{equation} \label{eq:photonCoalescenceSpectrum}
        \frac{\diff^2 n_a}{\diff t \, \diff\omega_a} \Bigg\rvert_{\gamma\gamma} = \left\lvert \gagDec \right\rvert^2 \frac{m_a^4}{128 \pi^3} \left(1 - \frac{4 m_\gamma^2}{m_a^2}\right) \, \Theta(m_a - 2 m_\gamma) \int_{\omega_\gamma^{{\rm min}}}^{\omega_{\gamma}^{{\rm max}}} \diff \omega_\gamma \, f_{\rm B}(\omega_\gamma) f_{\rm B}(\omega_a-\omega_\gamma) \, ,
    \end{equation}
    where $\omega_a$ is the ALP energy, $\omega_\gamma$ the energy of one of the photons, such that the other has energy $\omega_a - \omega_\gamma$, and the integration limits are as in the lepton fusion case
    \begin{equation}\label{eq:photonCoalescenceIntegrationLimits}
        \omega_\gamma^{\rm max,min} = \frac{\omega_a}{2} \pm \frac{\sqrt{(\omega_a^2 - m_a^2)(m_a^2 - 4 m_\gamma^2)}}{2 m_a} \, .
    \end{equation}
    The effective coupling for this process is the same as for its inverse, i.e., ALP-to-photon decay, and can be written as \cite{Ferreira:2022xlw,Bauer:2017ris}
    \begin{equation} \label{eq:effectiveDecayCoupling}
    \begin{aligned}
        \gagDec &= \frac{2\alpha}{\pi} \gal \left[ 1 + 2 (m_\ell^0)^2 C_0\left( m_\gamma^2, m_\gamma^2, m_a^2, (m_\ell^0)^2, (m_\ell^0)^2, (m_\ell^0)^2 \right) \right]\\
        &\simeq
        \frac{2\alpha}{\pi} \gal \times
        \begin{cases}
            - \frac{m_a^2 + 2 m_\gamma^2}{12 (m_\ell^0)^2} & \text{for } \max(m_a,m_\gamma) \ll m_\ell^0 \, ,\\
            1 & \text{for } \max(m_a,m_\gamma) \gg m_\ell^0 \, ,
        \end{cases}
    \end{aligned}
    \end{equation}
    i.e., it vanishes in the heavy-fermion limit and becomes constant for small $m_\ell^0$. Since the effective photon mass is typically larger than the vacuum electron mass, the coupling proportional to $\gae$ is independent of $m_a$ for most of the regions in the PNS in which photon coalescence is allowed kinematically. In the case of an ALP-muon coupling, both limits are relevant, since $m_\gamma \ll m_\mu$ but the ALP mass can be larger or smaller than the muon mass.

    The photon coalescence contribution is shown in yellow in \cref{fig:processComparison}.

    \subsection{Comparison} \label{subsec:comparison}
\begin{figure}[h]
        \centering
        \includegraphics[width=\linewidth]{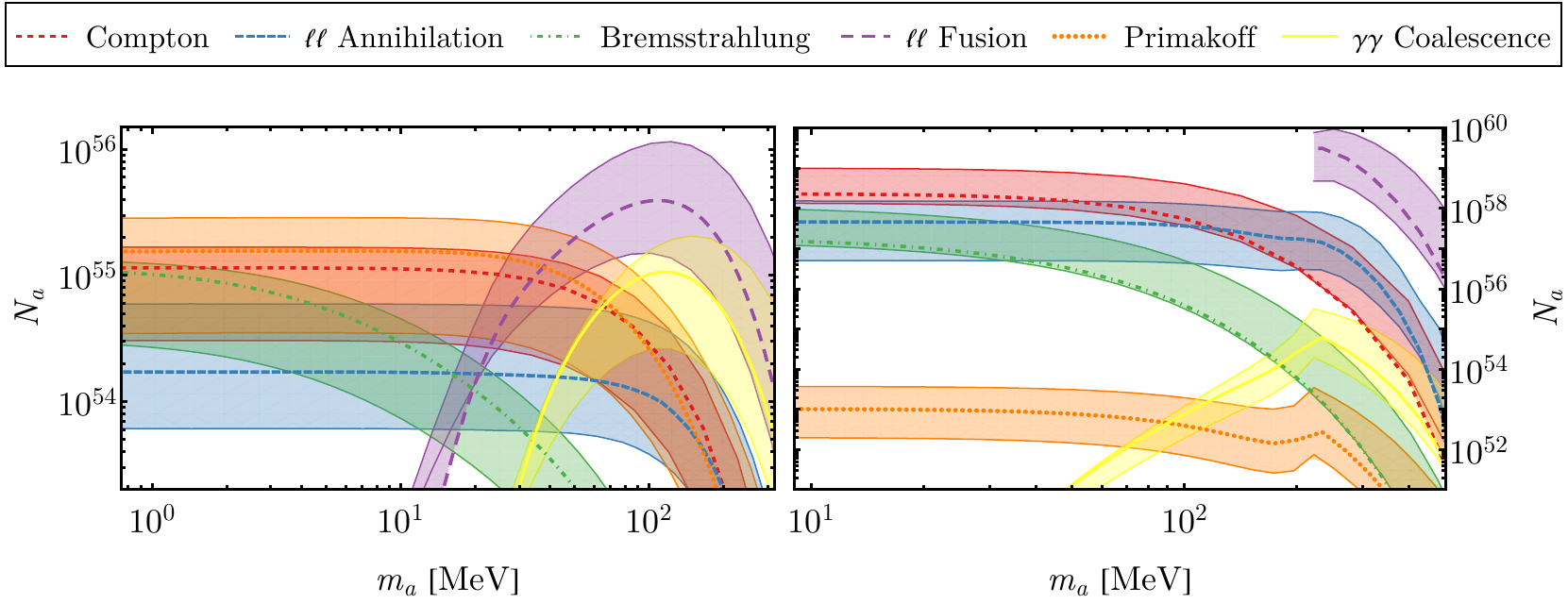}
        \caption{Comparison between the different contributions to ALP production in the SN models in the case of an ALP-electron coupling of $\gae = 10^{-10} \, \mathrm{MeV}^{-1}$ (left panel), and ALP-muon coupling of $\gamu = 10^{-10} \, \mathrm{MeV}^{-1}$ (right panel). The plots show the total number of produced ALPs, $N_a$, ignoring reabsorption and decay (see main text) as function of the ALP mass $m_a$. The transparent band shows the range of predictions from the three different SN models.
        }
        \label{fig:processComparison}
    \end{figure}

    In \cref{fig:processComparison} we compare the contributions to ALP production by the processes calculated in the previous sections. We show the ALP production spectrum as defined below in \cref{eq:alpSpectrum} integrated over radius, time, and energy, not taking into account that strongly coupled ALPs can also be reabsorbed (see \cref{subsec:cooling,subsec:calorimetricBound}). Thus, \cref{fig:processComparison} shows the number of ALPs that escape the SN in the free-streaming regime, i.e., when the couplings are small enough for (nearly) all particles to escape the stellar remnant.
    Thus, the shown ALP numbers simply scale as $\gal^2$ but do not necessarily represent any spectrum observable outside of the SN, for which reabsorption and decay could be relevant. When the couplings are too large, the dominant contributions to these total numbers from the inner-most hot and dense regions of the plasma are reabsorbed before leaving the PNS, and only the usually sub-dominantly contributing ALPs from the outer regions can escape.

    Different colors correspond to the different processes; for each process, we show the contribution in the case of the SN model DD2-18.0 in the line-style indicated in the legend, while the thinner solid lines in the respective color show the contributions in case of the Garching models, with the one corresponding to SFHo-18.8 always below the one for LS220-20.0 due to the higher temperatures and densities in the latter model. The transparent band indicates the spread between the highest and lowest value for $N_a$ in the three models, showing that the uncertainty in the SN model leads to about an order of magnitude uncertainty in the total number of ALPs produced.
    
    In the case of ALPs coupled to electrons, Compton, Primakoff, and bremsstrahlung processes contribute similarly for light ALPs with $m_a \lesssim 1 \, \mathrm{MeV}$. Bremsstrahlung is suppressed for all larger ALP masses \cite{Lucente:2020whw}, while Compton and Primakoff processes are Boltzmann suppressed when $m_a \gtrsim 40 \, \mathrm{MeV}$, around the typical peak temperature in the plasma. For larger masses, the inverse decay processes become kinematically possible and quickly exceed all other contributions for $m_a \gtrsim 40 \, \mathrm{MeV}$, with electron-positron fusion always more efficient than photon coalescence because the latter is a loop effect and is further phase-space suppressed by the effective photon mass being larger than the effective electron mass. The inverse decay processes are increasingly efficient over a range of ALP masses since the effective electron and photon masses are higher in hotter, denser regions of the SN plasma, allowing ALPs to be produced in progressively larger regions of the PNS with growing $m_a$.
 
    For ALPs coupled to muons, the picture is overall similar, however the loop-induced Primakoff and photon coalescence processes are significantly smaller than in the electron case. As discussed in \cref{subsec:primakoff,subsec:coalescence}, this is because for ALP masses and energies below the muon mass of 106 MeV, the effective coupling is suppressed according to \cref{eq:effectivePrimakoffCoupling,eq:effectiveDecayCoupling}, and additionally the tree-level couplings are enhanced as $\gamuScalar = 2 m_\mu^0 \gamu $. This also makes muon-anti-muon annihilation the dominant contribution around $m_a \lesssim 200 $~MeV, and even reaching about $10\%$ of the largest contribution (namely Compton) in the massless limit. As was estimated in Ref.~\cite{Caputo:2021rux}, we find that muon-nucleon Bremsstrahlung is subdominant in all SN models in the free-streaming case. Muon fusion is again the dominant process where it is kinematically allowed, i.e., for $m_a > 2 m_\mu$. Note that since the muon mass does not receive relevant thermal corrections, the inverse decay becomes suddenly available above the mass threshold, as opposed to the gradual transition in the electron case.

    Having computed the production spectra of all relevant processes, we can now use them to derive observational bounds on the leptophilic ALP parameter space in the following section.

\section{Observational bounds} \label{sec:bounds}
In this work, we consider a range of astrophysical phenomena on which the production of ALPs in ccSNe would have an observable effect. All of these effects have been studied before -- however, we improve here on the previous literature in several ways, among them by calculating all the contributions to the ALP production spectra discussed in the previous section to a high precision. 
In the following sections, we will shortly describe the respective observables and the resulting bounds, including specific improvements that we introduce here.

\begin{figure}
    \includegraphics[width=\linewidth]{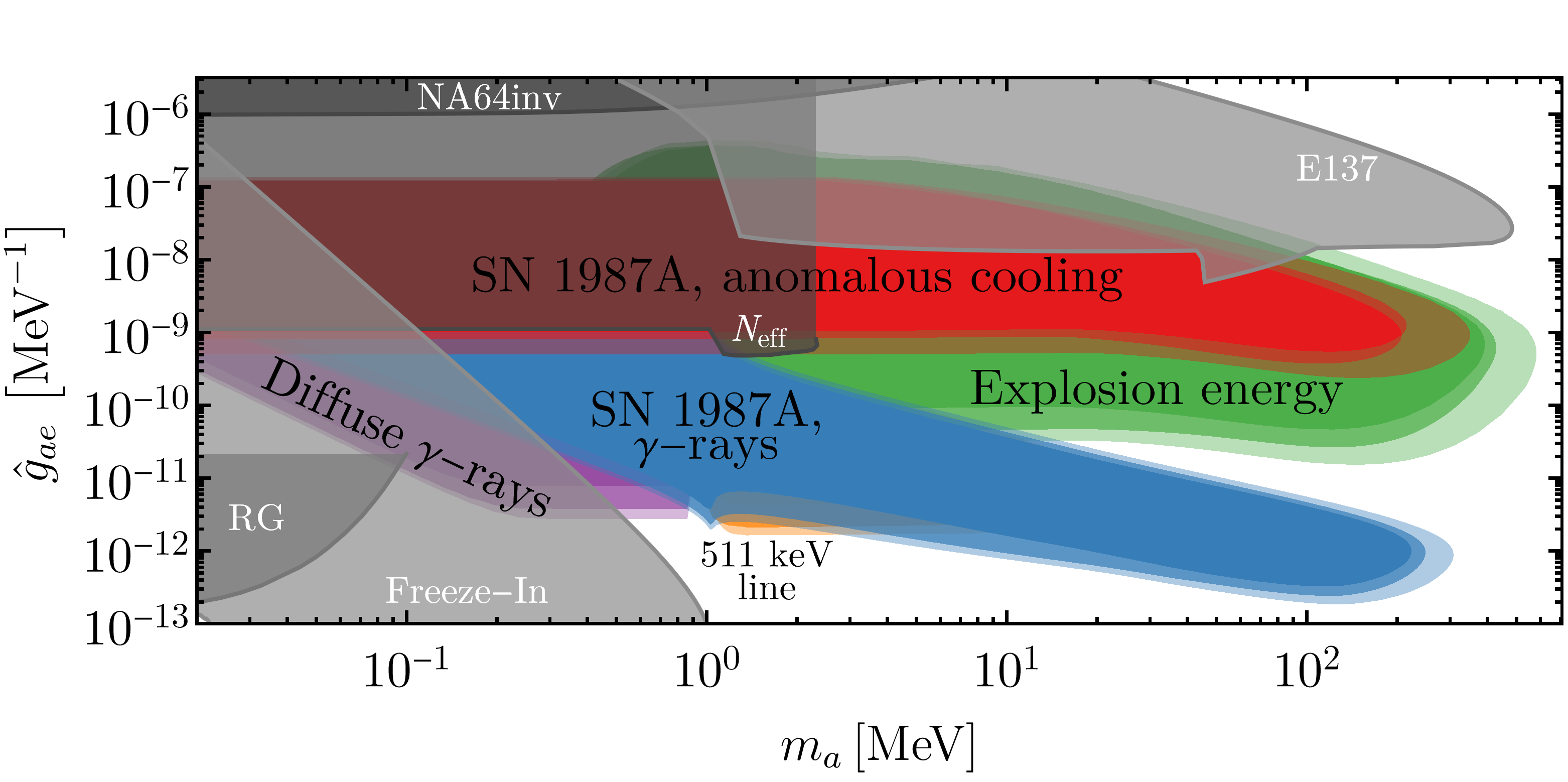}
    \\
    \includegraphics[width=\linewidth]{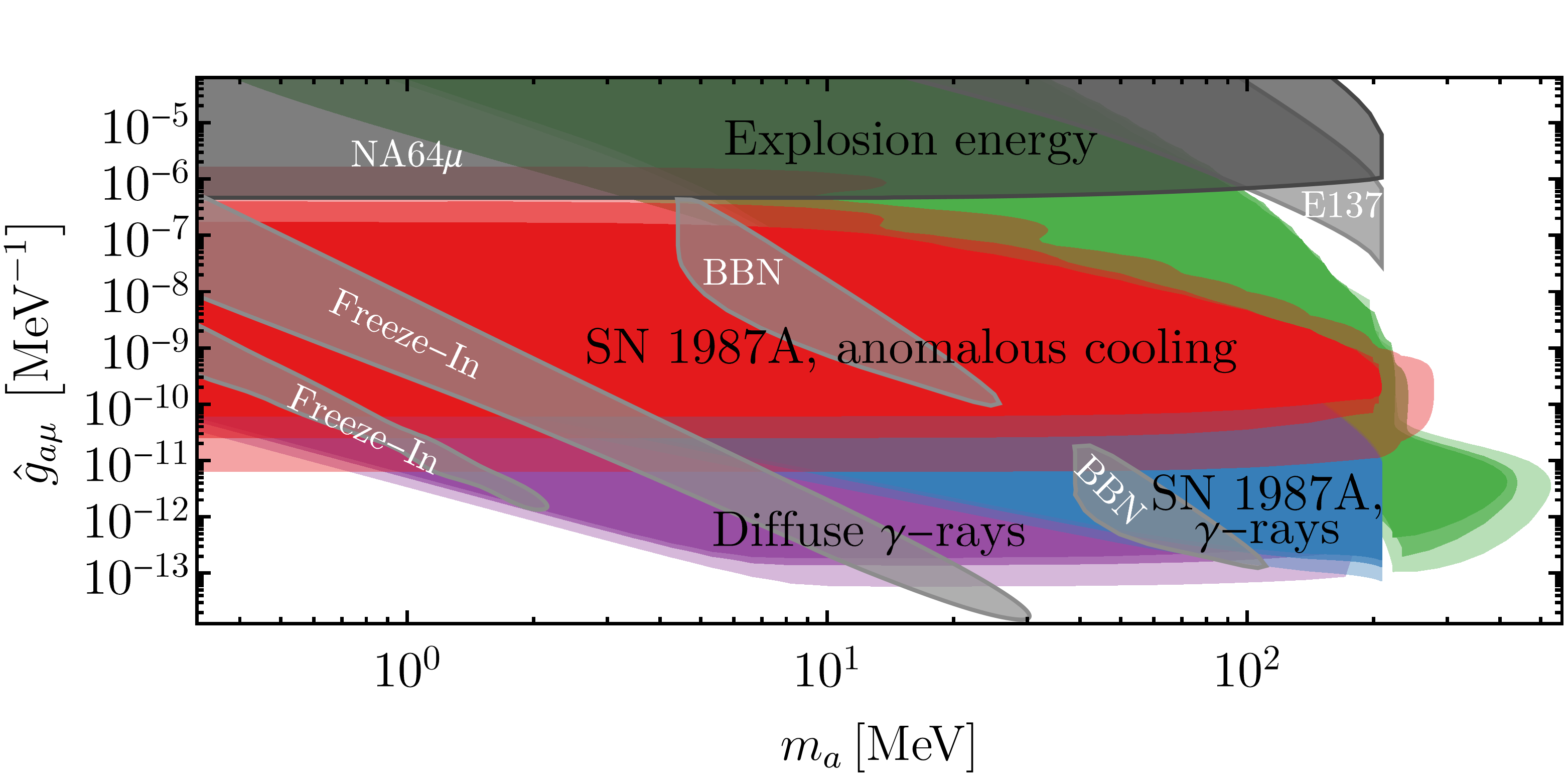}
    \caption{Bounds on the coupling of ALPs to electrons (upper panel) and muons (lower panel), respectively. We show the bounds from the three SN models employed here (see \cref{sec:SNmodel}) with different levels of opacity. Mostly, the largest region of each color corresponds to LS220-20.0, the smallest to SFHo-18.8, and the one in the middle to DD2-18.0. Details on each bound and how they are influenced by the different SN models can be found in \cref{subsec:cooling,subsec:decayBound,subsec:calorimetricBound,subsec:positronBound,subsec:diffuseGammaRayBound}.
    In shades of gray, we show other constraints that are relevant in this part of the ALP parameter space, as described in the main text.}
    \label{fig:exclusionPlots}
\end{figure}

We anticipate the summary of our results in \cref{fig:exclusionPlots}, where the excluded regions of the ALP parameter space are shown for ALPs coupled to electrons (upper panel) and muons (lower panel). We also show other relevant constraints in this region of the leptophilic ALP parameter space, namely, the constraints from the beam dump experiments E137 and NA64 (in two different modes, invisible and muon) \cite{Eberhart:2025lyu}, from cosmology via freeze-in production and its effect on the effective number of degrees of freedom ($N_{\rm eff}$), BBN\footnote{\label{footnote:cosmoConstraint}
        We have included the  BBN constraint ``islands'' on $\gamu$ but not the ones on $\gae$ that we showed in Ref.~\cite{Ferreira:2022xlw} because they were translated from the constraints on $\tau_{a \to \gamma \gamma}$ in Ref.~\cite{Depta:2020zbh}, which here is a loop-induced process, but the translation did not take into account the fact that for $m_a \geq 2 m_e$ the tree-level decays to electron and positrons will decrease the branching ratio of the decays to photons and shorten the total lifetime of the ALP, in some cases, to values below the time at which photo-disintegration of light nuclei becomes relevant. Therefore, to verify if there are leftover islands one would need to redo the analysis, including those two effects.}
and the X-ray and gamma-ray background \cite{Langhoff:2022bij,Ferreira:2022xlw,Depta:2020zbh}, and from red giants cooling data (RG) \cite{Carenza:2020zil}. 

    \subsection{Anomalous cooling of SN~1987A} \label{subsec:cooling}
    If ALPs would have been produced in SN~1987A that was observed 38 years ago in the Large Magellanic cloud, they would have constituted an additional cooling channel for the proto-neutron star (PNS) forming at the centre of the explosion. In the standard model, the PNS is cooled predominantly by the emission of neutrinos. Since the neutrino burst following SN~1987A was observed to last approximately 10 seconds, mostly in agreement with simulations \cite{Fiorillo:2023frv}, there is an upper limit on any additional, non-neutrino cooling of the PNS because the energy lost this way would cool the core faster and dynamically shorten the neutrino burst. As argued in Ref.~\cite{Raffelt:1990yz}, based on early simulations of the SN plasma \cite{Mayle:1987as,Mayle:1989yx,Burrows:1988ah}, the neutrino burst is not significantly shortened if the cooling power, or \emph{luminosity}, of ALPs, $L_a$, is below that of the neutrinos, $L_\nu$, one second after the shock of the explosion bounces in the SN core. To make this constraint more precise, state-of-the-art simulations of the SN explosion that include ALP emission (and, where relevant, also ALP absorption) are necessary, see e.g.~\cite{Fischer:2016cyd,Fischer:2021jfm,Betranhandy:2022bvr}. Here, we apply the simpler criterion $ L_a < L_\nu $ at $t_{\rm pb} = 1$~s to estimate the cooling bound on the ALP-electron and ALP-muon couplings.

    To calculate the luminosities, we use the three SN models described in \cref{sec:SNmodel}. 
    With SM-only simulations the production spectra of ALPs are calculated as functions of the post-bounce time $t_{\rm pb}$ and the position in the PNS plasma, which is fully specified by the radial distance $r$ from the centre of the explosion since we use spherically symmetric simulations. Because the plasma is in local thermal equilibrium, these coordinates determine all relevant thermodynamic quantities such as temperatures and densities.

    The neutrino luminosity is an output of the SN simulations, and in our case is $ L_\nu(t_{\rm pb} = 1 {\rm s}) = 3.0 \cdot 10^{52} \, {\rm erg}/{\rm s} $ for an observer at $r\to\infty$ for the Agile-Boltztran simulations \cite{Fischer:2021jfm}, $ L_\nu(t_{\rm pb} = 1 {\rm s}) = 4.4 \cdot 10^{52} \, {\rm erg}/{\rm s} $ for the SFHo-18.8 model, and $ L_\nu(t_{\rm pb} = 1 {\rm s}) = 8.3 \cdot 10^{52} \, {\rm erg}/{\rm s} $ for the LS220-20.0 model \cite{Caputo:2021rux}.
    The ALP luminosity, on the other hand, has to be calculated from the production spectra \cite{Caputo:2022rca}:
    \begin{equation}\label{eq:luminosityDefinition}
        L_a = \int_0^{R_\nu} \diff r \, 4\pi r^2 \, \lambda^2(r) \int_{m_a / \lambda(r)}^{\infty} \diff \omega_a \, \omega_a \, \frac{\diff^2 n_a}{\diff t \, \diff\omega_a}(r,\omega_a) \cdot \mathcal{T}(r,\omega_a) \, ,
    \end{equation}
    where all quantities are evaluated at $t_{\rm pb} = 1 $~s, $R_\nu$ is the radius of the neutrino sphere, in which the neutrino burst is primarily produced, $\lambda(r)$ is the lapse function, which accounts for general relativistic gravity corrections \cite{Caputo:2022mah} and which is tabulated in the simulation output as well, $\omega_a$ is the ALP energy in the frame of a local observer, and $\mathcal{T}(r, \omega_a)$ is the transmissivity. The transmissivity is the angular average of the probability that an ALP produced at radius $r$ with energy $\omega_a$ in a direction given by the angle $\theta$ between the radial direction and its momentum will escape  the PNS plasma. Following Ref.~\cite{Caputo:2022rca}, the transmissivity is given in terms of the optical depth, which in turn can be calculated from the production spectrum (related to what is called the reduced absorption rate in Ref.~\cite{Caputo:2022rca})\footnote{We neglected the quantum statistical factor $(1+f_a)$ when calculating the production spectrum, see \cref{eq:spectrumDefinition}. If we include it there, we need to cancel this factor in the optical depth, since the absorption probability of any given ALP does not depend on the overall density of ALPs. Thus, our expression for the optical depth is correct for all coupling strengths when using the approximated production spectrum with $f_a \to 0$ and since it is the transmissivity that determines the upper end of the constraint in the strong coupling regime (with only a mild logarithmic dependence on the production spectrum) our approximation is justified. \label{footnote:boseEnhancementProdSpectrum}}
    \begin{equation} \label{eq:opticalDepth}
    \begin{aligned}
        \mathcal{T}(r,\omega_a) &= \frac{1}{2} \int_{-1}^{1} \diff \cos \theta \, e^{-\tau(r, \omega_a, \cos\theta)} \, ,\\
        \text{with } \tau(r,\omega_a,\cos\theta) &= \frac{1}{2\pi^2} \int_0^{s_{\rm max}} \diff s \, \frac{\omega_a^2 - m_a^2}{\exp[\omega_a/T(r'(s))] - 1} \left[ \frac{\diff^2 n_a}{\diff t \, \diff\omega_a} \big( r'(s), \omega_a \big) \right]^{-1} \, ,\\
        \text{and } r'(s) &= \sqrt{r^2 + s^2 + 2 r s \cos\theta}\, , \,\, s_{\rm max} = \sqrt{R_{\rm far}^2 - (1-\cos^2\theta)r^2} - r \cos\theta \, ,
    \end{aligned}
    \end{equation}
    where $s$ is the distance the ALP has traveled since emission and $r'(s)$ is the radial coordinate along its path. We extend the integral to the finite radius $R_{\rm far}$, i.e., we integrate $s$ from $0$ to $s_{\rm max}$, since neutrinos can be efficiently produced at radii up to $R_{\rm far} > R_\nu$. We follow Ref.~\cite{Lucente:2020whw} and identify $R_{\rm far}$ with the neutrino gain radius, inside of which neutrino production is more efficient than neutrino absorption (i.e., there is a net energy gain from neutrinos). For the DDS2-18.0 model, we take $R_\nu = 21 \, \mathrm{km}$ and $R_{\rm far} = 24 \, \mathrm{km}$ \cite{Lucente:2021hbp}. For the models of the Garching group, we estimate the neutrino sphere radius from the criterion $\rho(R_\nu) = 10^{12} \mathrm{g} \, \mathrm{cm}^{-3}$, while the gain radii at all time steps are tabulated in the simulation output; we find $R_\nu = 19.1 \, \mathrm{km}$ and $R_{\rm far} = 21.1 \, \mathrm{km}$ for the SFHo-18.8 model, and $R_\nu = 18.2 \, \mathrm{km}$ and $R_{\rm far} = 19.7 \, \mathrm{km}$ for the LS220-20.0 model.
    
    By accounting for the possibility that energy deposited at radii between $R_\nu$ and $R_{\rm far}$ would be completely converted into neutrinos we make a conservative assumption, since probably not all the energy would be converted into neutrinos. Furthermore, this choice makes the numerical integration of \cref{eq:luminosityDefinition} feasible. If we were to account only for ALP absorption until $R_\nu$, the optical depth would exactly vanish at that radius, leading to an unbounded scaling $\frac{\diff L_a}{\diff r} \rvert_{R_\nu} \sim \gal^2$. If the fixed radial grid on which we evaluate the integrand in \cref{eq:luminosityDefinition} includes $R_\nu$, then $L_a$, obtained by quadrature, becomes arbitrarily large for increasing $\gal$. If the grid does not contain $R_\nu$, the resulting integral depends sensitively (and artificially) on the distance of the closest grid point to $R_\nu$. To accurately evaluate the integral in this case would require an adaptive grid with grid spacing on the size of the (ALP-parameter dependent) mean-free path, which would be very costly in practice. Additionally, in the recent paper Ref.~\cite{Fiorillo:2025sln},  the energy transfer contribution from ALPs emitted from radii larger than $R_\nu$ that propagate to smaller radii into the PNS and deposit energy there, which would be subtracted in \cref{eq:luminosityDefinition}, is taken into account. At large couplings, the positive and negative contributions nearly cancel, and a numerical approximation for this limit is employed in Ref.~\cite{Fiorillo:2025sln}. Also this approach alleviates the numerical difficulties that appear for a vanishing optical depth at $R_\nu$. In a more complete treatment, both contributions, backwards emitted ALPs as well as an estimate for losses due to neutrino production, should be included. Here, we choose to follow the approach of Ref.~\cite{Lucente:2020whw}.

    The definition of the optical depth in \cref{eq:opticalDepth} takes into account that ALPs are emitted isotropically in all directions at any given radius, of which those that travel in the radial outward direction have a shorter path through the colder, more dilute outwards regions of the SN core, while those traveling radially inwards have a much larger and ``optically'' dense path. In practice, we estimate the angular integral in \cref{eq:luminosityDefinition} by evaluating the integrand for nine evenly spaced values of $\cos\theta$, as a trade-off between accuracy and required computing resources.
    In the previous literature, the luminosity was mostly calculated without the angular average included in \cref{eq:luminosityDefinition}, instead the optical depth was either evaluated for straight, radially-outward trajectories only \cite{Chang:2018rso,Bollig:2020xdr,Ertas:2020xcc,Ferreira:2022xlw}, or a correction factor was applied \cite{Chang:2016ntp,Lucente:2020whw,Sung:2019xie} that in fact made the approximation even worse as pointed out in Refs.~\cite{Caputo:2021rux, Caputo:2022rca}. Thus, we improve on the previous literature not only by calculating all production spectra, without approximations such as ultrarelativistic ALPs or neglecting loop-level processes, but also by treating the absorption probabilities more realistically.

    The upper limit on anomalous cooling in SN~1987A leads to the exclusion of the red regions in the ALP parameter space shown in \cref{fig:exclusionPlots}, in the left (right) panel for an ALP coupled to electrons (muons). The largest excluded region corresponds to the LS220-20.0 model. In the other two cases, the regions overlap for the electron coupling, but the constraints from the SFHo-18.8 model are slightly offset towards larger couplings compared to the DD2-18.0 model. For a muonphilic ALP, the DD2-18.0 model yields the smallest excluded region.

    \subsection{Gamma-rays from SN~1987A} \label{subsec:decayBound}
    Relatively weakly coupled ALPs produced in the PNS can escape not only the plasma in the central region of the explosion, but even the stellar material that still surrounds the PNS 
    and that extends up to the progenitor star's radius $ R_* $. Subsequently, such ALPs can decay into a pair of gamma-ray photons which might be observed by gamma-ray telescopes in orbit around earth.

    We use the formalism developed in Refs.~\cite{Ferreira:2022xlw,Muller:2023vjm} to derive a bound from the non-observation of a gamma-ray burst following SN~1987A. The gamma-ray spectrometer of the Solar Maximum Mission was taking data for $\Delta t = 223$~s after the neutrino burst from the SN reached earth, and did not detect any excess over the background of gamma rays with energies between 25 to 100 MeV \cite{Chupp:1989kx,Oberauer:1993yr}. Thus, gamma-ray fluence is constrained to be $ F_\gamma < 1.78 \, \mathrm{cm}^{-2}$ \cite{Jaeckel:2017tud,Hoof:2022xbe}. Since the observation time is so short compared to the distance to the SN times the speed of light, we can use the ``small-angle'' approximation derived in Ref.~\cite{Muller:2023vjm} for the masses that we are interested in here, and therefore the ALP-induced gamma-ray fluence can be calculated as
    \begin{equation} \label{eq:fluence}
        F_\gamma = \int_{m_a}^{\infty} \diff \omega_a \int_{\omega_\gamma^{\rm min}}^{\omega_\gamma^{\rm max}} \diff \omega_\gamma \, \frac{{\rm BR}_{a \to \gamma\gamma}}{2\pi d_{\rm SN}^2} \, p_a^{-1} \, \frac{\diff N_a}{\diff \omega_a} \left[ \exp\left(-\frac{m_a R_*}{\tau_a p_a}\right) - \exp\left(-\frac{2 \omega_\gamma \, \Delta t}{\tau_a m_a}\right) \right] \Theta(\Delta\omega_\gamma) \, ,
    \end{equation}
    where $ R_* = 3 \times 10^{10} $~m is the radius of the progenitor of SN~1987A \cite{Kazanas:2014mca}, $ d_{\rm SN} = 51.4 $~kpc is its distance to earth \cite{Panagia:2003rt},
    and the integration region is given by \cite{Muller:2023vjm}
    \begin{equation}
    \begin{aligned}
        \omega_\gamma^{{\rm min}}(p_a) &= \max \left( 25 {\rm  MeV}, \, \frac{1}{2}(\omega_a - p_a), \, \frac{m_a^2 \, R_*}{2 \, p_a \, \Delta t} \right) \, , \\
        \omega_\gamma^{{\rm max}}(p_a) &= \min \left( 100 {\rm  MeV}, \, \frac{1}{2}(\omega_a + p_a) \right) \, , \\
        \Delta \omega_\gamma(p_a) &= \omega_\gamma^{{\rm max}}(p_a) - \omega_\gamma^{{\rm min}}(p_a) \, .
    \end{aligned}
    \end{equation}
    Furthermore, $\tau_a$ is the total lifetime of the ALP in its rest frame, i.e., the inverse of the sum of the decay rates in vacuum \cite{Bauer:2017ris,Ferreira:2022xlw}\footnote{Note that at the couplings relevant for the decay bound, absorption can be neglected and the ALPs always escape the SN plasma by assumption -- hence only the vacuum decay rates set the mean-free path of the ALPs in \cref{eq:fluence}.}
    \begin{equation} \label{eq:alpDecayRates}
    \begin{gathered}
        \tau_a^{-1} = \Gamma_{a\to\gamma\gamma} + \Gamma_{a\to\ell^{+}\ell^{-}} \, , \quad
        \Gamma_{a\to\gamma\gamma} = \frac{\left\lvert \gagDec \right\rvert^2 m_a^3}{64\pi} \, , \quad
        \Gamma_{a\to\ell^{+}\ell^{-}} = \frac{\gal^2 m_\ell^2}{2\pi} \sqrt{m_a^2 - 4 m_\ell^2} \,\Theta(m_a - 2 m_\ell) \, ,
    \end{gathered}
    \end{equation}
    and consequently, the branching ratio for the decay into a gamma-ray pair in the case of leptophilic ALPs is
    \begin{equation}
        \mathrm{BR}_{a\to\gamma\gamma} \equiv \frac{\Gamma_{a\to\gamma\gamma}}{\Gamma_{a\to\ell^{+}\ell^{-}} + \Gamma_{a\to\gamma\gamma}} \, ,
    \end{equation}
    with the effective decay coupling as defined in \cref{eq:effectiveDecayCoupling}. Note that the branching ratio is 1 for masses below the threshold $m_a < 2 m_\ell$ and that the effective mass and the vacuum mass agree outside of the SN plasma, which is the region in which the decays we are interested in occur, so that we can use here the lighter notation $m_\ell^0 \to m_\ell$ while for photons $m_\gamma \to 0$.
    
    The total number spectrum of ALPs produced in the SN, including gravitational corrections, is given by
    \begin{equation} \label{eq:alpSpectrum}
        \frac{\diff N_a}{\diff \omega_a} = 4\pi \int \diff t \int \diff r \, r^2 \, \lambda^{-1}(r,t) \dfrac{\diff^2 n_a}{\diff t \, \diff\omega_{a}^{\rm loc}}(r,t,\lambda^{-1}(r,t) \, \omega_a) \, ,
    \end{equation}
    where $\omega_a^{\rm loc}$ is the ALP-energy in the local frame, while $\omega_a = \lambda \, \omega_a^{\rm loc}$ is the energy outside of the progenitor star, and $\lambda(r,t)$ is the lapse factor accounting for relativistic corrections such as redshift and gravitational trapping \cite{Calore:2021hhn}. The production spectra calculated in \cref{sec:production} are differentials with respect to the local energy $ \omega_a^{\rm loc} $ even though we did not use that notation there. In practice, we integrate the radius $r$ in \cref{eq:alpSpectrum} up to $ 50 \, (100) \, \mathrm{km} $, and the time from $0.5 \, (0.25) \, \mathrm{s}$ to $13.3 \, (10) \, \mathrm{s}$ for the DDS2-18.0 (SFHo-18.8 and LS220-20.0) model, which captures the peak of the ALP production and yields numerically stable results. As in Refs.~\cite{Jaeckel:2017tud,Ferreira:2022xlw,Muller:2023vjm}, we approximate the ALP burst as instantaneous, i.e.~we do not take the different times of ALP production into account when we calculate the arrival time of the gamma rays for \cref{eq:fluence}, which was shown to be a good approximation in Ref.~\cite{Hoof:2022xbe} since the typical gamma-ray signal length is much longer than the $\sim 10$~s duration of ALP production.

    We note that, according to Ref.~\cite{Diamond:2023scc}, in some regions of the parameter space, all ALPs emitted from the SN might decay at large rates in a small enough volume such that the photon densities reached can create an electron-positron plasma, a so-called fireball. In this case, the gamma-ray photons would not be observable and the decay bound does not directly apply. However, the fireball loses its energy in the form of X-ray photons which would have been observed by the Pioneer Venus Orbiter Satellite (PVO), and therefore also these regions of parameter space are excluded \cite{Diamond:2023scc}. Since these regions are contained inside the decay bound that we derive here, we do not show them separately in \cref{fig:exclusionPlots}.

    Since \cref{eq:fluence} is a good approximation to the full expression for the fluence used in Ref.~\cite{Ferreira:2022xlw}, the only improvement for the decay bound on the ALP-electron coupling are the updated production spectra. For the ALP-muon coupling, our analysis is similar to Ref.~\cite{Caputo:2021rux} but we improve on the production spectra (which are only calculated approximately there) and we extend the mass range of the bound to heavier ALPs. Furthermore, Ref.~\cite{Caputo:2021rux} only derives bounds for the case of a pseudoscalar ALP-muon coupling, for which the loop-induced effective couplings to photons are different \cite{Ferreira:2022xlw}.
    
    The decay bound, resulting from $F_\gamma \leq 1.78 \, \mathrm{cm}^{-2}$, is shown in orange in \cref{fig:exclusionPlots}, in the left (right) panel in case of an ALP coupled to electrons (muons).

    Recently, it was shown \cite{Muller:2023pip} that the extragalactic SN 2023ixf might yield a comparable gamma-ray constraint on ALPs. However, since there is a larger uncertainty on the properties of the progenitor of SN 2023ixf, we do not include the bound in this work.
 
    \subsection{Explosion energy} \label{subsec:calorimetricBound}
    ALPs emitted from the SN core take energy away from the neutrino sphere and deposit this energy into the stellar plasma, if they decay in the mantle of the progenitor star. 
    The matter in the mantle will be ejected in the explosion. By depositing additional energy there, the ALPs contribute to the energy budget of this outburst, which for SNe with progenitor masses between 10 and 20 solar masses is observed to be around $10^{51} \, \mathrm{erg}$ \cite{Bruenn:2014qea,Nomoto:2013oal}. Thus, the additional energy deposition by exotic particles cannot exceed this observed value, as was originally pointed out for radiatively decaying neutrinos in Ref.~\cite{Falk:1978kf} and recently rediscovered for dark photons in Ref.~\cite{Sung:2019xie}.
    Even more recently, Ref.~\cite{Caputo:2022mah} argued that the bound is about an order of magnitude stronger because ccSNe with smaller explosion energies, of order $10^{50}$ erg (e.g.~\cite{Spiro_2014,Pumo:2016nsy}), are observed as well. However, since it is not clear that the ALP production spectra that we infer from the SN simulations we use (see \cref{sec:SNmodel}) are a good approximation for the production spectra in such low-energy SNe, we use the more conservative bound in this work.
    
    The energy that is deposited into the mantle of the SN's progenitor by ALPs from the centre of the explosion can be calculated as
    \begin{equation} \label{eq:explosionEnergy}
    \begin{aligned}
        E_{\rm mantle} = \int \diff t \int_0^{R_\nu} \diff r \int_{m_a / \lambda}^\infty \diff \omega_a \, 4\pi r^2 \lambda \, \omega_a \frac{\diff n_a}{\diff t \, \diff \omega_a}\Big(r, t, \omega_a\Big) \, \mathcal{T}(r, t, \omega_a)
        \left[ 1 - \exp\left(- \frac{R_* - r}{\ell_a(\lambda \, \omega_a)}\right) \right] \, ,
    \end{aligned}
    \end{equation}
    where we integrate $t$ as described below \cref{eq:alpSpectrum}, $ \lambda(r,t) $ is the lapse factor, $\ell_a = p_a \tau_a / m_a$ is the total ALP decay length in vacuum,\footnote{At radii comparable to $R_\star$ the mean-free path of the ALP is entirely dominated by decays since density and temperature at the outer parts of the progenitor star are far too low to lead to significant ALP absorption rates through scattering. Furthermore, the decay rate is the same as in vacuum, as in-medium effects such as thermal masses are negligible at these radii as well.} and $\mathcal{T}$ is the transmissivity as in \cref{eq:opticalDepth}. Without the transmissivity factor, this expression agrees with Ref.~\cite{Caputo:2022mah}.
    
    As in the case of the cooling bound, we evaluate the transmissivity by determining the probability of an ALP produced in the PNS to reach the larger radius $R_{\rm far}$, accounting for the possibility that energy deposited at smaller radii (between $R_\nu$ and $R_{\rm far}$) would be efficiently converted into neutrinos, which again would free-stream out of the stellar plasma and hence transport the energy they carry out of the star. As described in Ref.~\cite{Fiorillo:2025sln} and \cref{subsec:cooling}, energy transport by ALPs from outside the PNS inward should be taken into account as well, which in practice has an effect similar to separating $R_\nu > R_{\rm far}$ as in our definition of the optical depth. A more sophisticated post-processing ansatz would include both effects in some way. We note, however, that the potentially affected upper end of the explosion energy constraint is for many masses already ruled out by beam dump constraints.
    
    The resulting explosion energy bound is shown in green in \cref{fig:exclusionPlots}. As one would expect, it covers some of the parameter region between the cooling and the decay bound, which is where the ALPs are short-lived enough to decay before leaving the progenitor, hence they evade the decay bound, but are weakly coupled enough to evade the cooling bound. For the electron coupling, this region is large and we see that the explosion bound is around an order of magnitude stronger than the cooling bound. This is in agreement with the fact that the explosion energy is generated during a time period of around 10 seconds such that the maximal allowed average power is $10^{50} \, \mathrm{erg}/\mathrm{s}$, slightly more than two orders of magnitude below $L_\nu = 3 \cdot 10^{52} \, \mathrm{erg}/\mathrm{s} $. Since $L_a$ and $E_{\rm mantle}$ both scale as $\gae^2$, the resulting bounds are expected to be about an order of magnitude apart from one another, in agreement with our results.

    In the muon case, the decay bound is much closer to the cooling bound since below $m_a<2m_\mu$ only loop-induced decays into photons are possible, and thus the decay length is much longer. Therefore, the explosion energy constraint can only cover new parts of the parameter space for larger couplings or higher masses than the former two. We attribute the fact that it extends to stronger couplings than the cooling bound in part to the fact that the explosion energy is an integral over time. At early times, including around $t_{\rm pb} = 1$~s, when the cooling bound is evaluated, the muon abundance is larger at larger radii and does not vanish as rapidly with $r$ as it does at later times. Thus, reabsorption through the tree-level semi-compton scattering and $ a \to \mu\mu $ decay is only efficient at early times throughout the region of efficient ALP production. Later, a surface appears at which the muon density is high enough to produce a sizable number of ALPs but falls quickly enough for absorption (of ALPs on outward trajectories) to happen only for much larger couplings -- hence, the explosion energy bound extends to larger couplings than the cooling bound. For masses $m_a > 2 m_\mu$, muon fusion can very efficiently produce ALPs even at lower couplings, which are needed to keep the ALP lifetime long enough since now decays into muon anti-muon pairs become possible. In principle, such an additional region could have appeared in the cooling bound as well but the local maximum of the luminosity is not larger than $L_\nu$ in that case.

    Since many SM-only simulations tend to supply the propagating shock of the ccSN explosion with too little energy to revive it again after it stalls \cite{OConnor:2018sti,Burrows:2020qrp}, we note that ALPs could provide some contribution and assist the shock revival as pointed out in, e.g., refs.~\cite{Schramm:1981mk,Mori:2021pcv}. 
    However, in recent works \cite{Sung:2019xie,Caputo:2022mah} it has been argued that the ALP-parameters that are estimated to be necessary for an ALP-assisted shock revival are excluded by the explosion energy constraint. Clearly, assessing the dynamical impact of the ALPs on SN explosion requires an analysis that goes beyond the determination of the overall calorimetry (see, e.g., \cite{Takata:2025lyu} for recent work in this direction).

    \subsection{Diffuse gamma-ray background} \label{subsec:diffuseGammaRayBound}
    As discussed in \cref{subsec:decayBound}, the ALPs produced in a SN explosion can decay outside the stellar progenitor plasma and lead to an observable gamma-ray burst. Comparing the accumulated gamma-ray flux originating from ALPs from all SNe in the cosmic history with the measured, diffuse extragalactic background light yields an additional bound on the underlying ALP parameters \cite{Calore:2020tjw,Calore:2021klc}.

    To calculate this diffuse gamma-ray signal we solve the coupled Boltzmann equations in the expanding FLRW universe for the ALP and photon phase space densities, following Ref.~\cite{Fogli:2004gy,Kolb:1990vq}, but adapting their approach to massive ALPs. The  general derivation can be found in \Cref{app:diffuseGammaRayFlux} with the somewhat lengthy solution in \cref{eq:diffuseFluxGeneralSolution}. Here we only note that the ultrarelativistic limit of the diffuse photon flux is given by
    \begin{equation}
    \begin{split}
        \frac{\diff \Phi_\gamma^{\rm diff.}}{\diff \omega_\gamma}\Bigg|_{\rm UR} =
        2 \, {\rm BR}_{a \to \gamma\gamma} \int_{\omega_\gamma}^{\infty} \diff \omega_a \int_0^\infty \diff z
        & \left( 1 - e^{-\int_0^z \diff \tilde{z} \, H^{-1}(\tilde{z}) (1 + \tilde{z})^{-2} / \ell_a} \right)\\
        &\quad \times
        e^{-R_\star / [(1+z) \ell_a]} \frac{1+z}{\omega_a} \frac{\diff n_{\rm SN}}{\diff z}(z) \frac{\diff N_a}{\diff \omega_a}\Big((1+z) \omega_a\Big) \, ,
    \end{split} \label{eq:diffuseFluxURSolution}
    \end{equation}
    where $\frac{\diff n_{\rm SN}}{\diff z}(z)$ is the comoving number density of SNe per unit redshift, $\frac{\diff N_a}{\diff \omega_a}$ is the ALP spectrum emitted by a single SN, see \cref{eq:alpSpectrum}, $\ell_a$ is the ALP decay length, see below \cref{eq:explosionEnergy}, $R_\star$ is the typical radius of a SN progenitor's photosphere, and $H(z)$ is the Hubble parameter at redshift $z$. This ultrarelativistic limit is in agreement with the expression in Ref.~\cite{Caputo:2021rux}. We find relevant deviations from the ultrarelativistic limit for the heavier ALP masses that can be constrained in both, the electron and muon cases. Also, as remarked in \cref{app:diffuseGammaRayFlux}, the generalized expression in \cref{eq:diffuseFluxGeneralSolution} is of the same numerical complexity as the ultrarelativistic limit and can thus be used with comparable computing resources and time.

    Following Ref.~\cite{Calore:2020tjw}, we compare our predicted photon signal with measurements from two different detectors, depending on the energy. For 0.8 MeV $ < \omega_\gamma < 30 $~MeV, we use the diffuse gamma-ray background measurements of the COMPTEL satellite that can be fitted as \cite{Kappadath:1998aa}
    \begin{equation}
        \frac{\diff \Phi_\gamma^{\rm COMPTEL}}{\diff \omega_\gamma} = 1.32 \times 10^{-3} \left(\frac{\omega_\gamma}{5 \, {\rm MeV}}\right)^{-2.4} {\rm cm}^{-2} \, {\rm s}^{-1} \, {\rm MeV}^{-1} \, ,
    \end{equation}
    while for larger energies, $ \omega_\gamma > 30 $~MeV, the relevant observations by \textit{Fermi}-LAT can be fitted as \cite{Calore:2020tjw}
    \begin{equation}
        \frac{\diff \Phi_\gamma^{\rm LAT}}{\diff \omega_\gamma} = 5.06 \times 10^{-6} \left(\frac{\omega_\gamma}{50 \, {\rm MeV}}\right)^{-2.2} {\rm cm}^{-2} \, {\rm s}^{-1} \, {\rm MeV}^{-1} \, .
    \end{equation}
    Note that we state the (angle-)integrated fluxes here, which, for the isotropic background, are just $4\pi$ times the often quoted differential flux values.
    Comparing the predicted contribution from SN ALPs for all energies with these values yields the diffuse gamma-ray bounds shown in purple in \cref{fig:exclusionPlots} ---  i.e., the constraint is at each point the strongest across all energy bins.
    It excludes ALPs with masses $m_a < 2m_\ell$ and relatively low couplings. The branching ratio for decays into photons decreases suddenly at the threshold $m_a > 2m_\ell$, and since already below the threshold the ALPs decay mostly promptly (on cosmological scales), this reduction in the branching ratio directly reduces the diffuse gamma-ray flux so much that no new regions are excluded at higher masses.

    \subsection{511 keV line} \label{subsec:positronBound}
    It is expected that core-collapse supernovae explode in our own Galaxy at a rate $\Gamma_{cc} \simeq 2 /$century \cite{Rozwadowska:2020nab,DelaTorreLuque:2023huu} and the cumulative ALP emission of such explosions would also lead to further observable signals.    
    
    In this section, we follow Ref.~\cite{DelaTorreLuque:2023huu} and use the measurement of the flux of 511 keV photons in our galaxy by the SPI instrument \cite{Bouchet:2010dj,Siegert:2015knp} to constrain ALPs from ccSNe. 
    In the presence of a $\gae$ coupling, the ALP can directly decay into electrons and positrons. But also for an ALP coupled to muons a sizeable number of positrons could be produced by the ALP decaying into a muon and an anti-muon, which subsequently decays almost immediately (on astronomical scales) into a pair of neutrinos and a positron. In any case, the produced positrons travel distances  $\lambda_{e^+} \sim 1 $~kpc, depending on their injection energy \cite{Calore:2021lih}, until they annihilate almost at rest into two 511 keV photons via the formation of a para-positronium state. The positrons can also annihilate in-flight, with non-vanishing kinetic energies, yielding a similarly strong constraint \cite{Balaji:2025alr} that we do not consider further here.
    
    The contribution to the photon line flux per unit of solid angle is \cite{Calore:2021klc,DelaTorreLuque:2023huu} 
    \begin{equation}
        \frac{\diff \Phi_\gamma^{511 \, {\rm keV}}}{\diff\Omega} = 2 k_{\rm ps} N_{\rm pos} \Gamma_{\rm SN} \int \diff s \, \frac{n_{\rm SN}(s, \Omega)}{4\pi}
    \end{equation}
    where $k_{\rm ps}=1/4$ is the estimated fraction of positrons that annihilate via para-positronium, $n_{\rm SN}$ is the volume distribution of SNe in the Milky Way and thus is not isotropic as opposed to the cosmological SN explosion rate of the previous section and $s$ is distance from the SN to the sun. The factor of two in the previous equation accounts for the two photons emitted in the process.
    Finally, the number of injected positrons, $N_{\rm pos}$, is related to the time-integrated flux of ALPs produced in the explosions of galactic ccSNe, $\diff N_a/\diff \omega_a$, that escaped the SN envelope and that decay into positrons within our galaxy via
    \begin{equation} \label{eq:positronNumberGalaxy}
    \begin{aligned}
        N_{\rm pos}
        &= \int \diff \omega_a \, {\rm BR}_{a \to e^{+}e^{-}} \frac{\diff N_a}{\diff \omega_a} \left[\exp(- R_* / \ell_a) - \exp(- R_{\rm Gal}/\ell_a) \right]\\
        &\simeq \int \diff \omega_a \, {\rm BR}_{a \to e^{+}e^{-}} \frac{\diff N_a}{\diff \omega_a} \exp(- R_* / \ell_a) \left[ 1 - \exp(- R_{\rm Gal}/\ell_a) \right]
    \end{aligned}
    \end{equation}
    where $\ell_a=\gamma_a \beta_a \tau_a$ is the \emph{total} decay length of the ALP.
    The second line is the expression given in Ref.~\cite{Calore:2021lih,Calore:2021klc}, which is an excellent approximation since $R_{\rm Gal} = 1~\mathrm{kpc} \gg R_*$
    
    In Ref.~\cite{DelaTorreLuque:2023huu} the SPI observations and two different models of the distribution of SNe in the galaxy, $n_{\rm SN}$, were used to constrain the number of positrons created in a single SN. Due to the uncertainties on the modelling of $n_{\rm SN}$ and the diffusion length and time scales of positrons in the Galaxy, the resulting bound on $ N_{\rm pos} $ can vary up to an order of magnitude. Here, we take the strongest constraint derived in Ref.~\cite{DelaTorreLuque:2023huu} because, as we will see, even in this case the resulting bound on $\gae$ is almost completely superseded by the decay bound. Hence, we impose that the ALP-electron coupling does not cause $N_{\rm pos}$ to exceed $1.7 \times 10^{52}$. The small excluded parameter region is shown in orange in \cref{fig:exclusionPlots}.

    That almost no new regions of the $ (m_a, \gae) $ parameter space can be excluded by this bound confirms the estimate of Ref.~\cite{Calore:2021klc} that ALP-production through the electron coupling alone does not yield a relevant contribution to the 511 keV line (in that reference, an additional coupling to nucleons is therefore assumed). 
    Here, we go beyond Ref.~\cite{Calore:2021klc} by including loop-induced interactions and the Compton and pair annihilation processes in the production spectrum. However, we find that the conclusions of Ref.~\cite{Calore:2021klc} remain robust.
    
    Since anti-muons decay into positrons with a branching ratio very close to 1, and their decay length is negligible compared to the diffusion length of positrons in the Galaxy \cite{ParticleDataGroup:2022pth}, \cref{eq:positronNumberGalaxy} also applies to ALPs coupled to muons with the replacement $ {\rm BR}_{a \to e^{+}e^{-}} \to {\rm BR}_{a \to \mu^{+}\mu^{-}} $. However, since ${\rm BR}_{a \to \mu^{+}\mu^{-}}$ vanishes below the threshold $ m_a = 2 m_\mu $ and the decay length of ALPs for such large masses is only larger than $R_*$ if $\gamu \lesssim 10^{-14} {\rm MeV}^{-1} $, there are never enough ALPs outside the SN that could decay into positrons via anti-muons to leave an observable signal.

\section{Conclusion and Outlook} \label{sec:conclusion}
In this work, we provide a comprehensive analysis  of the  phenomenology of massive ALPs that are produced in ccSN explosions via couplings to electrons or muons. The central element of our study is the determination of the ALP production spectra, as detailed in \cref{sec:production}. Our calculation incorporates contributions via the Compton process, lepton-pair annihilation, lepton-proton bremsstrahlung, lepton fusion, the Primakoff process, and photon coalescence.  

For the ALP-electron coupling the inclusion of the full Compton process is an important improvement upon our earlier work \cite{Ferreira:2022xlw}. For the ALP-muon coupling, we confirm the estimate of Ref.~\cite{Caputo:2021rux} that the Compton process and muon fusion are the most relevant contributions to ALP production. We show that this conclusion is robust also for massive ALPs.

From these spectra, we derive a set of constraints based on observations of individual, Galactic, and extragalactic ccSNe, as discussed in  \cref{sec:bounds}. These results exclude regions of parameter space spanning nearly seven orders of magnitude in the couplings $\gae$ and $\gamu$, and extending up to $ m_a \sim 500$~MeV. These constraints are summarized in \cref{fig:exclusionPlots}. In addition to refining the production spectra, we introduce several methodological improvements relative to previous studies:  for the cooling bound, we include non-radial trajectories into our determination of the luminosity; for the explosion energy bound, we add absorption effects that are important for large couplings, at the upper end of the bound; and for the diffuse gamma-ray background we extend the literature beyond the ultrarelativistic limit.

This work is a contribution to the study of the effective field theory of ALPs, and can be extended by including all ALP-Standard Model interactions (potentially) relevant in the SN plasma: a photon coupling (which, as discussed in \cref{subsec:leptonCouplings}, also corresponds to pseudoscalar lepton couplings) and couplings to baryons or mesons. Finally, also a coupling to the tau-lepton might be relevant, mostly through the loop-induced interaction with photons (cf.~\cref{subsec:effectivePhotonCoupling}) since due to their large mass the thermal population of tau leptons in the SN plasma is strongly suppressed.

Our findings further establish core-collapse supernovae as uniquely sensitive probes of physics beyond the Standard Model. At the same time, by employing three independent ccSN models — characterized by distinct equations of state, numerical implementations, and astrophysical parameters — we systematically assess some of the theoretical uncertainties intrinsic to this class of studies. Such an assessment is essential for the robust interpretation of parameter exclusions and underscores the need for further investigations incorporating an even broader range of supernova models and astrophysical inputs.

\section*{Acknowledgements}
The authors would like to thank Tobias Fischer and Thomas Janka for sharing their SN simulation data with us and very helpful discussions on the topic.
This research utilized the Sunrise HPC facility supported by the Technical Division at the Department of Physics, Stockholm University. Furthermore, part of the computation done for this project was performed on the UCloud interactive HPC system, which is managed by the eScience Center at the University of Southern Denmark.
RZF acknowledges the financial support provided by FCT - Fundação para a Ciência e Tecnologia, I.P., through  
the Strategic Funding UID/04650/2025 and 
national funds with DOI identifiers 10.54499/2023.11681.PEX, and 10.54499/2024.00249.CERN funded by measure RE-C06-i06.m02 – ``Reinforcement of funding for International Partnerships in Science, Technology and Innovation'' of the Recovery and Resilience Plan - RRP.
DM gratefully acknowledges support from the Swedish Research Council (VR) under grants 2019-02337 and 2024-04289. ER acknowledges partial funding from Villum Experiment grant no.~00028137 as well as support from Manuel Meyer through guidance and mentorship. This article is based upon work from COST Action COSMIC WISPers CA21106, supported by COST (European Cooperation in Science and Technology).

\appendix
\section{Photon flux from massive ALP decays throughout cosmic history}
\label{app:diffuseGammaRayFlux}
In this appendix, we derive the diffuse gamma-ray flux from SN-ALP decays by solving the Boltzmann equations for the phase space densities of ALPs and photons in the expanding universe. Both densities are homogeneous and isotropic since the cosmological SN rate is homogeneous and the ALP emission from any SN is assumed to be isotropic, simplifying the Boltzmann equations significantly. First, we can define rescaled comoving phase-space densities for massive ALPs and photons as follows:
\begin{equation}
    \tilde{f}_a(p_a,t) = 4\pi \, p_a^2 \left(\frac{a(t)}{a_0}\right)^3 f_a(p_a,t) \, , \qquad \tilde{f}_\gamma(\omega_\gamma,t) = 4\pi \, \omega_\gamma^2 \left(\frac{a(t)}{a_0}\right)^3 f_\gamma(\omega_\gamma,t) \, ,
\end{equation}
where $a(t)$ is the cosmological scale factor, $a_0$ is its value today, i.e.~at time $t_0$, the $f_i$ are the usual phase-space densities, and $p_a$ and $\omega_\gamma$ are the ALP momentum and photon energy, respectively. With these definitions, the Boltzmann equations will simplify and the number densities today become just one-dimensional momentum integrals over the $\tilde{f}_i$ evaluated at the present time $t_0$:
\begin{equation}
    n_i^0 = \int_0^\infty \, \diff p_i \tilde{f}_i(p_i, t_0) \, .
\end{equation}
Plugging the definitions above into the Boltzmann equation in an expanding Universe yields \cite{Kolb:1990vq,Fogli:2004gy}
\begin{equation}
\begin{aligned}
    \left[\partial_t - H(t) p_a \partial_{p_a} - H(t)\right] \tilde{f}_a(p_a, t) &= r_{\rm SN}(t) \frac{\diff N_a}{\diff p_a}
    (p_a) - \frac{m_a}{\omega_a} \Gamma_{a, {\rm tot.}} \tilde{f}_a(p_a, t) \, ,\\
    \left[\partial_t - H(t) \omega_\gamma \partial_{\omega_\gamma} - H(t)\right] \tilde{f}_\gamma(\omega_\gamma, t) &= \int_{\left\lvert \omega_\gamma - \frac{m_a^2}{4 \omega_\gamma} \right\rvert}^\infty \, \diff p_a \, p_a^{-1} \cdot 2 \, \frac{m_a}{\omega_a}\Gamma_{a \to \gamma\gamma} \cdot \tilde{f}_a(p_a, t) \, ,
\end{aligned}
\end{equation}
where $H(t) = \dot{a}(t) / a(t)$ is the Hubble parameter, which, written as a function of cosmological redshift $z = a(t)^{-1} - 1$ and in the redshift range of interest to us, can be evaluated as $H(z) = H_0 \sqrt{\Omega_\Lambda + (1+z^3) \Omega_m}$, and we take the approximate observed values $\Omega_m \simeq 0.3, \, \Omega_\Lambda \simeq 0.7, \, H_0 \simeq 2.2 \cdot 10^{-22} \, {\rm s}^{-1}$ \cite{Caputo:2021rux}.
The first equation, describing the evolution of the ALP density, contains a source term due to the emission of ALPs from SNe, where $r_{\rm SN}$ is the homogeneous SN rate, i.e., the number of SN explosions per time and volume, and $\frac{\diff N_a}{\diff p_a} = \frac{p_a}{\omega_a} \frac{\diff N_a}{\diff \omega_a}$ is the time-integrated momentum-differential ALP spectrum emitted by a single SN, including the effect of gravitational redshift from the ALP leaving the SN, as in \cref{eq:alpSpectrum}. Furthermore, there is a sink term for the ALP density due to their decays to photons and leptons, where $\Gamma_{a, {\rm tot.}} = \Gamma_{a \to \gamma\gamma} + \Gamma_{a \to \ell^{+}\ell^{-}}$ is the total rest-frame decay rate of the ALP, cf.~\cref{eq:alpDecayRates}, which is divided by a time-dilation factor since the ALPs are moving.
In the second equation, describing the evolution of the photon density, there is only a source term, due to the decay $a \to \gamma\gamma$. The decay rate is integrated over the probability density for an ALP of momentum $p_a$ to decay into a photon of energy $\omega_\gamma$ (and another photon of energy $\omega_a - \omega_\gamma$, which is taken into account by the factor of 2 in front of the decay rate), which is simply $p_a^{-1}$ \cite{Muller:2023vjm}.

Adapting the general solution of Ref.~\cite{Fogli:2004gy} to the case of massive ALPs, we find for the differential diffuse photon flux today (which, in natural units, is equal to $\tilde{f}_\gamma(\omega_\gamma, t_0)$):
\begin{equation} \label{eq:diffuseFluxTripleIntegral}
\begin{split}
    \frac{\diff \Phi_\gamma^{\rm diff.}}{\diff \omega_\gamma} = 2 \, {\rm BR}_{a \to \gamma\gamma}\int_0^\infty \diff z \int_z^\infty \diff z' \int_{\hat{p}_a^{\rm min.}(z)}^\infty \diff \hat{p}_a \, \frac{(1+z')^2}{(1+z) H(z)} \left[\frac{m_a^2}{\hat{p}_a^2} + (1+z)^2\right]^{-1/2} \left[\frac{m_a^2}{\hat{p}_a^2} + (1+z')^2\right]^{-1/2}\\
    \times \frac{\diff n_{\rm SN}}{\diff z}(z') \, \frac{\diff N_a}{\diff \omega_a}\Big((1+z') \hat{p}_a\Big) \frac{m_a}{\hat{p}_a^2} \Gamma_{a, {\rm tot.}} \exp\left[- \frac{m_a}{\hat{p}_a} \Gamma_{a, {\rm tot.}} \left(\xi_a(z') - \xi_a(z)\right)\right] \, ,
\end{split}
\end{equation}
where
\begin{equation}
    \hat{p}_a^{\rm min.}(z) = \left\lvert \omega_\gamma - \frac{m_a^2}{4 \, (1+z)^2 \, \omega_\gamma} \right\rvert \, , \qquad
    \xi_a(z^{(\prime)}) = \int_0^{z^{(\prime)}} \diff \tilde{z} \, H^{-1}(\tilde{z}) (1 + \tilde{z})^{-1} \left[\frac{m_a^2}{\hat{p}_a^2} + (1+\tilde{z})^2\right]^{-1/2} \, ,
\end{equation}
such that the argument of the exponential in \cref{eq:diffuseFluxTripleIntegral} is just a time-integral over the redshifted inverse life-time of the ALP.

The gamma-ray flux is thus an integral over the redshift at which a population of SNe exploded, $z'$, the redshift at which the ALPs decay into photons, $z$, and the momentum of the ALP at the time of decay, $\hat{p}_a$.
The function $ \xi_a(z) $ is strictly increasing and asymptotes to a finite value $ \xi_a(z \to \infty) = \xi_a^\infty $, and hence we can switch integration variables, yielding:
\begin{equation}
\begin{aligned}
    \frac{\diff \Phi_\gamma^{\rm diff.}}{\diff \omega_\gamma} &= 2 {\rm BR}_{a \to \gamma\gamma} \int_0^\infty \diff z' \int_0^\infty \diff \hat{p}_a \frac{(1+z')^2}{\hat{p}_a} \left[\frac{m_a^2}{\hat{p}_a^2} + (1+z')^2\right]^{-1/2} \frac{\diff n_{\rm SN}}{\diff z}(z') \frac{\diff N_a}{\diff \omega_a}\Big((1+z') \hat{p}_a\Big)\\
    &\, \times \int_0^{\xi_a^\infty} \diff \xi_a
    \left(\frac{\partial}{\partial \xi_a} \exp\left[- \frac{m_a}{\hat{p}_a} \Gamma_{a, {\rm tot.}} \left(\xi_a(z') - \xi_a\right)\right] \right) \Theta\left(\hat{p}_a - \left\lvert \omega_\gamma - \frac{m_a^2}{4 \, [1+z(\xi_a)]^2 \, \omega_\gamma} \right\rvert\right) \Theta(z' - z(\xi_a)) \, ,
\end{aligned}
\end{equation}
where $z(\xi_a)$, which only appears in the argument of the Heaviside functions, does not have a closed-form analytical expression. Reformulating the first Heaviside function, we can write
\begin{equation}
\begin{split}
    \Theta\left(\hat{p}_a - \left\lvert \omega_\gamma - \frac{m_a^2}{4 \, (1+z)^2 \, \omega_\gamma} \right\rvert\right)
    = \Theta\left(z - z_{+}\right)
    \left[\Theta(\hat{p}_a - \omega_\gamma) + \Theta(\omega_\gamma - \hat{p}_a) \Theta\left(z_{-} - z\right) \right] \, , \\
    \text{with } z_{\pm} = \frac{m_a}{2 \sqrt{\omega_\gamma(\omega_\gamma \pm \hat{p}_a)}} - 1 \, ,
\end{split}
\end{equation}
where $ z_{+} < z_{-} $ but neither is necessarily positive and thus neither is necessarily contained in the integration region.
With this form of the Heaviside function we can partially integrate the integral in the second line, use the distributional derivative $\Theta'(x) = \delta(x)$ and the fact that the ALP spectrum vanishes exponentially for $z \to \infty$, and rename the integration variable $z' \to z$. Thus, we find
\begin{equation} \label{eq:diffuseFluxGeneralSolution}
\begin{aligned}
    &\frac{\diff \Phi_\gamma^{\rm diff.}}{\diff \omega_\gamma} = \Bigg\{
    \int_{\max\left(\omega_\gamma,\left\lvert  \omega_\gamma - \frac{m_a^2}{4 \omega_\gamma}\right\rvert\right)}^{\infty} \diff \hat{p}_a \int_0^\infty \diff z \left( 1 - e^{-\tilde\Gamma \xi_a} \right)\\
    &+ \Theta\left(\frac{m_a}{2\sqrt{2}} - \omega_\gamma\right) \int_{\omega_\gamma}^{\frac{m_a^2}{4 \omega_\gamma} - \omega_\gamma} \diff \hat{p}_a \int_{z_{+}}^\infty \diff z \left( 1 - e^{-\tilde\Gamma (\xi_a - \xi_a^{+})} \right)\\
    &+ \Theta\left(\frac{m_a}{2} - \omega_\gamma\right) \int_0^{\min\left(\omega_\gamma, \frac{m_a^2}{4 \omega_\gamma} - \omega_\gamma\right)} \diff \hat{p}_a \left[ 
        \int_{z_{+}}^{z_{-}} \diff z \left( 1 - e^{-\tilde\Gamma (\xi_a - \xi_a^{+})} \right)    
        + \int_{z_{-}}^\infty \diff z \, e^{-\tilde\Gamma \xi_a} \left( e^{\tilde\Gamma \xi_a^{-}} - e^{\tilde\Gamma \xi_a^{+}} \right)
    \right]\\
    &+ \Theta\left(\omega_\gamma - \frac{m_a}{2\sqrt{2}}\right) \int_{\left\lvert  \omega_\gamma - \frac{m_a^2}{4 \omega_\gamma}\right\rvert}^{\omega_\gamma} \diff \hat{p}_a \left[ 
        \int_{0}^{z_{-}} \diff z \left( 1 - e^{-\tilde\Gamma \xi_a} \right)    
        + \int_{z_{-}}^\infty \diff z \, e^{-\tilde\Gamma \xi_a} \left( e^{\tilde\Gamma \xi_a^{-}} - 1 \right)
    \right] \Bigg\}\\
    &\hspace{9cm} \times
    2 \, {\rm BR}_{a \to \gamma\gamma} \frac{1+z}{\hat{p}_a} \frac{\diff n_{\rm SN}}{\diff z}(z) \frac{\diff N_a}{\diff \omega_a}\Big((1+z) \hat{p}_a\Big) \, ,
\end{aligned}
\end{equation}
with
\begin{equation}
    \tilde{\Gamma} = \frac{m_a}{\hat{p}_a} \Gamma_{a, {\rm tot.}}, \quad \xi_a \equiv \xi_a(z), \quad \xi_a^{\pm} \equiv \xi_a(z_{\pm}) \, ,
\end{equation}
where the expression in curly brackets should be understood as an integral operator acting on the common integrand in the last line of \cref{eq:diffuseFluxGeneralSolution}. We arranged the integrands, including the exponentials in the curly brackets, in such a way that they are all manifestly positive on their respective integration regions, to facilitate numerically stable integration.
In the ultrarelativistic limit, $m_a \to 0$, the above reduces to the simpler expression found in Ref.~\cite{Caputo:2021rux} and \cref{eq:diffuseFluxURSolution}, since only the integral in the first line contributes. However, even in the present more general case with a non-zero mass the integral is only 2-dimensional and can be carried out numerically relatively easily if the function $\xi_a(z, m_a / \hat{p}_a)$ is tabulated and interpolated. Note that the flux found in Ref.~\cite{Calore:2020tjw} is not the ultra-relativistic limit of the expression found here and indeed also disagrees with Ref.~\cite{Caputo:2021rux}.

\bibliography{SNAxionMuonElectron}
\bibliographystyle{JHEP}
	
\end{document}